\documentclass[a4paper,11pt]{article}
\usepackage{jheppub} 
\usepackage{lineno}
\usepackage{cancel}
\def\mathcolor#1#{\@mathcolor{#1}}
\def\@mathcolor#1#2#3{%
	\protect\leavevmode
	\begingroup\color#1{#2}#3\endgroup
}

\arxivnumber{2312.01879} 
\title{\boldmath$\mathcal{N}=2$ conformal supergravity in five dimensions}

\author{Soumya Adhikari \& Bindusar Sahoo}
\affiliation{School of Physics, Indian Institute of Science Education and Research Thiruvananthapuram, \\
Vithura, Thiruvananthapuram, India}
\emailAdd{soumya12physics20@iisertvm.ac.in, bsahoo@iisertvm.ac.in}

\abstract{ $\mathcal{N}=2$ conformal supergravity in five dimensions is constructed via a systematic off-shell reduction scheme from maximal conformal supergravity in six dimensions which is $(2,0)$. The dimensional reduction of the $(2,0)$ Weyl multiplet in six dimensions gives us the Weyl multiplet in five dimensions which is a dilaton Weyl multiplet as it has a dilaton scalar. The dimensional reduction of the $(2,0)$ tensor multiplet in six dimensions gives us the $\mathcal{N}=2$ vector multiplet in five dimensions coupled to conformal supergravity. We also comment on Nahm's classification regarding the non-existence of an $\mathcal{N}=2$ superconformal algebra in five dimensions and why it does not contradict the existence of $\mathcal{N}=2$ conformal supergravity in five dimensions that is constructed in this paper.}

\begin{document}
	\maketitle
	\flushbottom
 \vspace{-1cm}
	\section{Introduction}
	\label{sec:intro}
	
	Conformal supergravity \cite{Kaku:1978nz} is a supersymmetric generalization of conformal gravity. It is invariant under bosonic gauge symmetries such as diffeomorphism, local Lorentz transformations (M), dilatation (D), special conformal transformations (K) and R-symmetry as well as fermionic gauge symmetries such as an ordinary supersymmetry (Q-susy) and a special supersymmetry (S-susy). As one can see, conformal supergravity has some extra symmetries such as dilatation, special conformal transformation, R-symmetry as well as S-supersymmetry that are not present in a physical matter-coupled supergravity theory. However, it turns out that conformal supergravity is a useful theoretical framework for the construction of such physical matter-coupled supergravity theories \cite{Ferrara:1978rk,Kaku:1978ea}.  The advantages of the framework of conformal supergravity in constructing matter coupled supergravity theories are as follows. Firstly it has an off-shell formulation and secondly because of its high degree of symmetries, the physical degrees of freedom needed for the construction of a matter coupled supergravity theory sits inside shorter multiplets. Both these properties allow for the construction of a matter coupled supergravity theory in a tractable manner.
	
	The standard way of formulating conformal supergravity is to start with a gauge theory corresponding to the superconformal algebra $SU(2,2|\mathcal{N})$ \cite{Ferrara:1977ij,Ferber:1976sp}. In order to convert the gauge theory into a theory of gravity, certain conventional constraints are imposed on the gauge theory curvatures/field strengths. The curvature constraints render some of the gauge fields as dependent. The bosonic and fermionic degrees of freedom of the remaining independent gauge fields do not match. As a result of which some additional auxiliary matter fields need to be systematically included while simultaneously modifying the supersymmetry transformation rules and the superconformal algebra \cite{Bergshoeff:1980is}. This whole procedure goes by the name of superconformal tensor calculus which is a component field approach\footnote{The superspace-superfield \cite{Gates:1983nr,Buchbinder:1998qv,Howe:1980sy,Howe:1981gz,Butter:2011sr,Butter:2012xg,Butter:2014gha,Butter:2009cp,Butter:2014xxa,Howe:1981ev,Kuzenko:2022skv,Kuzenko:2022ajd} approach offers a systematic alternative to superconformal tensor calculus.}. The resulting multiplet of the gauge fields together with the auxiliary matter fields is known as the Weyl multiplet. It contains the graviton and its supersymmetric partner gravitino. In the modified superconformal algebra, the structure constants are replaced by structure functions which depend on the auxiliary matter fields that are introduced for the completion of the multiplet. Such an algebra is known as soft algebra \cite{Freedman:2012zz}.  
	
	Apart from the Weyl multiplet, the next most important ingredients in constructing a physical matter coupled supergravity theory are the matter multiplets. The matter multiplets can be of two types: physical and compensating. The physical matter multiplets ultimately couple to the physical supergravity theory whereas the compensating matter multiplets are used as compensators to gauge fix the additional symmetries of the conformal supergravity theory such as dilatation, special conformal transformation, R-symmetries and S-susy so that one can obtain the physical supergravity theory where these additional symmetries are not present, also known as Poincar{\'e} supergravity theory \cite{VanProeyen:1983wk,deRoo:1984zyh}. 
	
	The Weyl multiplet comes in different variants depending on the way the auxiliary matter fields are chosen. There is a standard way of choosing the auxiliary matter fields which leads to what is known as the ``standard Weyl multiplet'' \cite{Bergshoeff:1980is,Bergshoeff:1999db}. However, there is a different way to chose the auxiliary matter fields. This choice involves a scalar field that transforms non-trivially under dilatation, also known as the dilaton. Because of which such a variant Weyl multiplet is also known as the ``dilaton Weyl multiplet''. Such variant Weyl multiplets were discovered in six dimensions for $(1,0)$ conformal supergravity \cite{Bergshoeff:1985mz}, in five dimensions for $\mathcal{N}=1$ conformal supergravity \cite {Bergshoeff:2001hc,Fujita:2001kv} and in four dimensions for $\mathcal{N}=2$ conformal supergravity \cite{Butter:2017pbp}. Sometimes the dilaton Weyl multiplet can also come in different variant. One such example, known as hyper-dilaton Weyl multiplet has been studied in four dimensions for $\mathcal{N}=2$ conformal supergravity \cite{Gold:2022bdk}.
	
	Some crucial features of the dilaton Weyl multiplet are as follows:
	\begin{itemize}
		\item 
		The dilaton Weyl multiplet itself consists of fields which directly compensates for the additional symmetries present in conformal supergravity theory such as dilatation, special conformal transformation and S-supersymmetry. And hence one should expect that the flat rigid limit of the soft superconformal algebra realized on the dilaton Weyl multiplet should be a super-Poincar{\'e} algebra instead of a superconformal algebra. On the other hand the standard Weyl multiplet does not come with any compensating field inside the multiplet. And hence the flat rigid limit of the soft superconformal algebra realized on the standard Weyl multiplet would be a superconformal algebra instead of a super-Poincar{\'e} algebra. 
		\item
		As a consequence of what is discussed above, one would need fewer compensating matter multiplets to go from conformal supergravity theory to Poincar{\'e} supergravity theory as compared to a similar construction using the standard Weyl multiplet. This feature makes the dilaton Weyl multiplet more suitable for the construction of a physical matter-coupled supergravity theory. 
		\item In order to construct supersymmetrization of curvature square terms in supergravity theories, both the Weyl multiplets can be used. However as seen in the literature, due to some mapping between the dilaton Weyl multiplet and the Yang-Mills multiplet, the dilaton Weyl multiplet is more useful in constructing the Riemann square term, which can then be used to obtain the supersymmetrization of arbitrary curvature squared invariants.~\cite{Ozkan:2013nwa,Mishra:2020jlc}\footnote{See \cite{Gold:2023ymc,Gold:2023ykx} for construction of curvature squared invariants in five dimensional minimal gauged supergravity and \cite{Gold:2023dfe,Butter:2014xxa} for construction of curvature squared invariants in five dimensional minimal supergravity from superspace.}.
	\end{itemize}
	The maximal supersymmetry for which conformal supergravity is possible in six dimensions is $(2,0)$. The $(2,0)$ superconformal algebra in six dimensions has 32 supercharges encoded in two symplectic Majorana right chiral and left chiral $Q$ and $S$ supercharges. One would expect that upon dimensional reduction on a circle one would obtain a conformal supergravity theory in five dimensions with 32 supercharges. In five dimensions, an irreducible spinor is a symplectic Majorana spinor which encodes 8 fermionic degrees of freedom. And hence the 32 supercharges in a conformal supergravity theory in five dimensions would be distributed between two symplectic Majorana Q-supercharges and two symplectic Majorana S-supercharges. Hence we will refer to such a theory as $\mathcal{N}=2$ conformal supergravity in five dimensions\footnote{A symplectic Majorana spinor in five dimensions is given in terms of a pair of spinor $\psi^{i}$ (where $i=1,2$) satisfying the $SU(2)$-Majorana condition. And hence two symplectic Majorana spinors will be given in terms of a quartet $\psi^{i}$ (where $i=1,2,3,4$) satisfying a $USp(4)$-Majorana condition. We will discuss this in detail in the appendix-\ref{conv}. There are some papers in the literature which refers to the theory of conformal supergravity with one pair of symplectic Majorana Q and S-supercharges as $\mathcal{N}=2$ \cite{Bergshoeff:2001hc}. However we will refer to such theories as $\mathcal{N}=1$ as opposed to what we are discussing which we will refer to as $\mathcal{N}=2$.}. Please refer to Appendix-\ref{conv} for the definition of a symplectic-Majorana spinor in five dimensions and symplectic-Majorana Weyl spinor in six dimensions. 
	
	As per a classification by Nahm \cite{Nahm:1977tg}, there is no rigid $\mathcal{N}=2$ superconformal algebra in five dimensions. However as discussed earlier, a theory of conformal supergravity is realized on soft superconformal algebra instead of a rigid superconformal algebra. The rigid limit of the soft superconformal algebra may or may not lead to a rigid superconformal algebra depending on what is the underlying Weyl multiplet i.e. dilaton or standard. The non existence of an $\mathcal{N}=2$ rigid superconformal algebra in five dimensions is a hint to the fact that there cannot be a standard Weyl multiplet in $\mathcal{N}=2$ conformal supergravity in five dimensions. And hence one would expect that the dimensional reduction of a $(2,0)$ standard Weyl multiplet in six dimensions on a circle will lead to a dilaton Weyl multiplet in five dimensions. We will see this explicitly.
	
	The rest of the paper is organized as follows: In section-2, we discuss six-dimensional $(2, 0)$ standard Weyl multiplet as well as the tensor multiplet in the conformal supergravity background following the reference \cite{Bergshoeff:1999db}. In section-3, we give the details of the systematic off-shell dimensional reduction scheme from six dimensions to five dimensions. In section-4, we present the $\mathcal{N}=2$ dilaton Weyl multiplet in five dimensions which we get from the six dimensional $(2, 0)$ standard Weyl multiplet using the dimensional reduction scheme discussed in section-3. We also give the transformation laws of all the components of the multiplet along with the soft superconformal algebra which is realized on this multiplet. In section-5, we dimensionally reduce the six dimensional $(2,0)$ tensor multiplet in the conformal supergravity background to a five dimensional $\mathcal{N}=2$ vector multiplet in the conformal supergravity background. In section-6, we conclude along with some discussions and future directions.  
\section{Six Dimensional $(2,0)$ Standard Weyl Multiplet \& the Tensor Multiplet}	
The $(2,0)$ conformal supergravity in six dimensions is based on the superconformal algebra $OSp(8^{*}|4)$. The standard Weyl multiplet for $(2,0)$ conformal supergravity in six dimensions that describes $128+128$ off-shell degrees of freedom is given in table \ref{table: 1}.

 \begin{table}[t!]
		\centering
		\centering
		\begin{tabular}{ |p{1cm}|p{7cm}|p{2.5cm}|p{2.5cm}| }
			\hline
			Field&Properties&USp(4) irreps&Weyl weight\\
			\hline
			\multicolumn{4}{|c|}{Independent Gauge fields}\\
			\hline
			$e_M{}^A$&vielbein &\bf{1}&-1\\
			$\psi_M^i$&Symplectic Majorana Weyl,$\Gamma_*\psi_M^i=\psi_M^i$\;, gravitino &\bf{4}&-$\frac{1}{2}$\\
			$V_M^{ij}$&$V_M^{ij}=V_M^{ji}$\;, $USp(4)$-gauge field &\bf{10}&0\\
			\hline
			\multicolumn{4}{|c|}{Dependent Gauge fields}\\
			\hline
			$\omega_{M}{}^{AB}$& $\omega_{M}^{AB}=-\omega_{M}^{BA}$, Spin Connection&\bf{1}&0\\
			$\phi_{M}^i$&Symplectic Majorana Weyl,$ \Gamma_* \phi_{M}^i=-\phi_{M}^i$&\bf{4}&$\frac{1}{2}$\\
			$f_M{}^A$& Special Conformal Gauge Field&\bf{1}&1\\
			\hline
			\multicolumn{4}{|c|}{Auxiliary matter fields}\\
			\hline
			$T^{ij}_{ABC}$  & Totally Anti-Symmetric, $\Omega_{ij}T^{ij}_{ABC}=0, T^{ij}_{ABC}=-\frac{1}{6}\epsilon_{ABCDEF}T^{ij}_{DEF}$&\bf{5}&1\\
			$D^{ij,kl}$& $D^{ij,kl}=-D^{ji,kl}=-D^{ij,lk},D^{ij,kl}=D^{kl,ij}, \Omega_{ij}D^{ij,kl}=0,{\Omega_{ik}\Omega_{jl}D^{ij,kl}=0}$&\bf{14}&2\\
			$\chi_k{}^{ij}$ &Symplectic Majorana Weyl, $\chi_k{}^{ij}=-\chi_k{}^{ji}, \Omega_{ij}\chi_k{}^{ij}=0, \chi_i{}^{ij}=0, \Gamma_* \chi_k{}^{ij}=+\chi_k{}^{ij}$&\bf{16}&$\frac{3}{2}$\\
			\hline
		\end{tabular}
		\caption{Six Dimensional Standard Weyl Multiplet}
		\label{table: 1}	
	\end{table}

	Here the $A,B,..$ and $M,N,..$ indices are the local Lorentz indices and the world indices respectively. The world indices run from $0$ to $5$ and the local Lorentz indices run from $\underline{0}$ to $\underline{5}$. The $i,j,..$ indices are the USp(4) indices that runs form 1 to 4. All the spinors mentioned above in six dimensions are symplectic Majorana Weyl with the chirality constraint as mentioned in the table. The  gauge fields $\omega_M^{AB}, f_M^A, \phi_M^i$ are dependent due to the curvature constraints as shown below. 
	\begin{align}\label{curv_const}
		R(P)_{MN}{}^A&=0\nonumber \\
		R(M)_{MN}{}^{AB}e^N{}_{B}+\frac{1}{4}T_{MBC}^{ij}T^{ABC}_{ij}&=0\nonumber\\
		\Gamma^M R(Q)_{MN}{}^i&=0\;,
	\end{align} 
	
	where the curvatures appearing above are the fully supercovariant curvatures for local translation (P), local Lorentz transformation (M), and Q-supersymmetry (Q) and are defined in (\ref{sup_cov_curv}) along with the other curvatures. All the gauge transformations of the independent fields are given below, where $\epsilon$, $\eta$, $\Lambda$, $\Lambda_A$, $\Lambda^{A}{}_{B}$, $\Lambda^{ij}$ denotes the parameters of Q-supersymmetry, S-supersymmetry, dilatation, special conformal transformation, local Lorentz transformation and $USp(4)$ R-symmetry respectively:
	{\allowdisplaybreaks
		\begin{align}\label{6dgaugetransf}
			\delta e_M{}^A&=-\Lambda_D e_M{}^A-\Lambda^{AB}e_{MB}+\frac{1}{2}\bar{\epsilon}\Gamma^A \psi_M\nonumber\\
			\delta \psi_M^i&=-\frac{1}{2}\Lambda_D\psi_M^i+\frac{1}{2}\Lambda^i{}_j\psi_M^j-\frac{1}{4}\Lambda^{AB}\Gamma_{AB}\psi_M^i+\mathcal{D}_M\epsilon^i+\frac{1}{24}T_{ABC}^{ij}\Gamma^{ABC}\Gamma_M\epsilon_j+\Gamma_M\eta^i\nonumber\\
			{\delta}b_M&={\partial_M\Lambda_D-2e_M^A\Lambda_{A}}-\frac{1}{2}\bar{\epsilon}\phi_M+\frac{1}{2}\bar{\eta}\psi_M\nonumber\\
			{\delta} V_M^{ij}&=\partial_M \Lambda^{ij}+\Lambda^{(i}_{~k}V^{j)k}_M-4\bar{\epsilon}^{(i}\phi^{j)}_M-\frac{4}{15}\bar{\epsilon}_k\gamma_M\chi^{(i,j)k}-4\bar{\eta}^{(i}\psi_M^{j)}\nonumber\\
			{\delta}T^{ij}_{ABC}&=\Lambda_D T_{ABC}^{ij}+3T^{ij}_{F[AB}\Lambda^{F}{}_{C]}-\Lambda^{[i}{}_{k}T_{ABC}^{j]k}+\frac{1}{8}\bar{\epsilon}^{[i}\Gamma^{DE}\Gamma_{ABC}R_{DE}^{j]}(Q)-\frac{1}{15}\bar{\epsilon}^k\Gamma_{ABC}\chi_k{}^{ij}\nonumber\\
			&-\Omega\text{trace}\nonumber\\
			\delta{\chi}_k{}^{ij}&=+\frac{3}{2}\Lambda_D \chi_{k}{}^{ij}-\Lambda^{[i}{}_{l}\chi_{k}{}^{j]l}-\frac{1}{2}\chi_l{}^{ij}\Lambda^{l}{}_{k}-\frac{1}{4}\Lambda^{AB}\Gamma_{AB}\chi_k{}^{ij}+\frac{5}{32} D_M \Gamma \cdot T^{ij}\Gamma^M\epsilon_k\nonumber\\
			&-\frac{15}{16}\Gamma\cdot R_k^{~[i}(V)\epsilon^{j]}-\frac{1}{4}D_{kl}^{ij}\epsilon^l+\frac{5}{8}\Gamma\cdot T^{ij}\eta_k-\text{traces}\nonumber\\
			{\delta}D^{ij,kl}&=\Lambda_D D^{ij,kl}-\Lambda^{[i}{}_{m}D^{j]m,kl}-2\bar{\epsilon}^{[i}\cancel{D}\chi^{j],kl}+4\bar{\eta}^{[i}\chi^{j],kl}+(ij \leftrightarrow kl)-\Omega\text{trace}\;, \nonumber\\
	\end{align}}

	where \begin{align}
		\mathcal{D}_M\epsilon^i=\partial_M \epsilon^i +\frac{1}{2}b_M\epsilon^i+\frac{1}{4}\omega_M{}^{AB}\Gamma_{AB}\epsilon^i-\frac{1}{2}V_M{}^i{}_j \epsilon^j.
	\end{align}
	In the transformations above as well as later in the paper, there are expressions where we subtract $\Omega$ trace and $\delta$ trace by simply writing $-\Omega$ trace and $-\delta$ trace. Whenever we write $-$ traces, this would mean subtracting both $\Omega$-trace as well as $\delta$-trace. This subtraction of traces should be clear from the $USp(4)$ representation of the field whose transformation is being given. 

The expressions for the fully super-covariant curvatures are as follows which can be obtained from the gauge transformations of the corresponding gauge fields given in (\ref{6dgaugetransf}, \ref{6ddep}):

	{\allowdisplaybreaks
		\begin{align}\label{sup_cov_curv}
			R(P)_{MN}{}^{A}&=2\partial_{[M}e_{N]}^A+2b_{[M}e_{N]}^A+{2\omega_{[M}^{~AB}e_{N]B}}-\frac{1}{2}\bar{\psi}_M\Gamma^A\psi_N\nonumber\\	
			R(M)_{MN}{}^{AB}&={2\partial_{[M}\omega_{N]}{}^{AB}}+2\omega_{[M}{}^{AC}\omega_{N]C}{}^{B}{-8f_{[M}^{~[A}e_{N]}^{~B]}}+\bar{\psi}_{[M}\Gamma^{AB}\phi_{N]}\nonumber\\
			&+\bar{\psi}_{[M}\gamma^{[A}R(Q)_{N]}^{~B]}+\frac{1}{2}\bar{\psi}_{[M}\Gamma_{N]}R(Q)^{AB}+\frac{1}{2}\bar{\psi}_{M,i}\gamma_C\psi_{N,j}T^{ABC,ij}\nonumber\\
			{R}(Q)_{MN}{}^i&=2\partial_{[M}\psi_{N]}^i+b_{[M}\psi_{N]}^i+{\frac{1}{2}\Gamma\cdot\omega_{[M}\psi_{N]}^i}-V^i_{~j[M}\psi_{N]}^j{+2\Gamma_{[M}\phi_{N]}^i}\nonumber\\
			&+\frac{1}{12}T^{ij}\cdot\Gamma\Gamma_{[M}\psi_{N]j}\nonumber\\
			R(V)_{MN}{}^{ij}&=2\partial_{[M}V_{N]}{}^{ij}+V_{[M}{}^{k(i}V_{N]}{}^{j)}{}_k+8\bar{\psi}_{[M}{}^{(i}\phi_{N]}{}^{j)}+\frac{8}{15}\bar{\psi}_{[M,k}\Gamma_{N]}\chi^{(i,j)k}\nonumber \\
		\end{align}
	}
	
The expressions for the dependent fields can be obtained by solving the curvature constraints given in (\ref{curv_const}):	
	\begin{equation}
		\begin{aligned}
			\omega_M{}^{AB}&=2e^{N[A}\partial_{[M}e_{N]}^{~B]}-e^{P[A}e^{B]Q}e_M^{~C}\partial_P e_{QC}+2e_M^{~[A}b^{B]}+\frac{1}{2}\bar{\psi}_M \gamma^{[A}\psi^{B]}+\frac{1}{4}\bar{\psi}^A\Gamma_M\psi^B\\
			f_M^{~A}&=-\frac{1}{8}R^{'}(M)_M^{A}+\frac{1}{80}e_M^{~A}R^{'}(M)+\frac{1}{32}T_{MCD}^{ij}T_{ij}^{ACD}\\
			\phi_M^i&=\frac{1}{16}\bigg(\gamma^{AB}\Gamma_M-\frac{3}{5}\Gamma_M\Gamma^{AB}\bigg)R^{'}(Q)_{AB}{}^i
		\end{aligned}  
	\end{equation}
	In the above expressions, we have defined the following:
	{\allowdisplaybreaks
		\begin{align}
			R^{'}(M)_{MN}{}^{AB}&=R(M)_{MN}{}^{AB}|_{f=0}\nonumber \\
			R^{'}(Q)_{MN}^{i}&=R(Q)_{MN}^{i}|_{\phi=0} \nonumber \\
			R^{'}(M)_{M}^{A}&=e^{N}_{B}R^{'}(M)_{MN}{}^{AB}\;, R^{'}(M)=e^{M}_{A}R^{'}(M)_{M}^{A}
	\end{align}}

	\subsection{The (2,0) Tensor Multiplet}\label{6dtensor}
	The matter multiplet in $(2,0)$ conformal supergravity in six dimensions is a tensor multiplet which is an on-shell multiplet describing $8+8$ on-shell degrees of freedom. The components of this multiplet are given in table-\ref{table:2}.
	
	\begin{table}[t!]
		\centering
		\centering
		\begin{tabular}{ |p{1.5cm}|p{1.5cm}|p{7cm}|p{1cm}|p{1cm}|}
			\hline
			Field & Type& Properties & USp(4)&Weyl weight\\
			\hline
			$B_{MN}$& Boson& real antisymmetric tensor gauge field, self-dual Field strength&\bf{1}&0\\
			$\psi^i$&Fermion&Symplectic Majorana, $\Gamma_*\psi^i=-\psi^i$&\bf{4}&$\frac{5}{2}$\\
			$\phi^{ij}$&Boson&$\phi^{ij}=-\phi^{ji}, \Omega_{ij}\phi^{ij}=0$&\bf{5}&2\\
			\hline
		\end{tabular}
		\caption{The (2,0) Tensor Multiplet}
		\label{table:2}	
	\end{table}

	The complete gauge transformations of the $(2,0)$ tensor multiplet in a conformal supergravity background are given by:
	
	\begin{equation}\label{Tensor}
		\begin{aligned}
			{\delta}B_{MN}&=-\bar{\epsilon}\Gamma_{MN}\psi+\bar{\epsilon}^i\Gamma_{[M}\psi_{N]}^j\phi_{ij},\\
			{\delta}\psi^i&=\frac{5}{2}\Lambda_D\psi^i+\frac{1}{2}\Lambda^i{}_j\psi^j-\frac{1}{4}\Lambda^{AB}\Gamma_{AB}\psi^i+\frac{1}{48}H_{MNP}^{+}\Gamma^{MNP}\epsilon^i+\frac{1}{4}\cancel{D}\phi^{ij}\epsilon_j-\phi^{ij}\eta_j,\\
			{\delta}\phi^{ij}&=2\Lambda_D\phi^{ij}-\Lambda^{[i}{}_{k}\phi^{j]k}-4\bar{\epsilon}^{[i}\psi^{j]}-\text{trace}
		\end{aligned}
	\end{equation}

	The field equations of the (2,0) tensor multiplet are given below:
	
	\begin{align}\label{EOM}
		H^-_{ABC}-\frac{1}{2}\phi_{ij}T^{ij}_{ABC}&=0,\nonumber\\
		\cancel{{D}}\psi^i-\frac{1}{15}\phi^{kl}\chi^{i}{}_{kl}-\frac{1}{12}T^{ij}_{ABC}\Gamma^{ABC}\psi_j&=0,\nonumber\\
		{D}^A{D}_A\phi_{ij}-\frac{1}{15}D^{kl}{}_{ij}\phi_{kl}+\frac{1}{3}H^+_{ABC}T^{ABC}_{ij}+\frac{16}{15}\bar{\chi}^k{}_{ij}\psi_k&=0.\nonumber\\
	\end{align}
	
	where \begin{align}
		{D}_M\psi^i&=\bigg(\partial_M-\frac{5}{2}b_M+\frac{1}{4}\omega_M{}^{AB}\Gamma_{AB}\bigg)\psi^i -\frac{1}{2}V_M^i{}_j\psi^j-\frac{1}{48}H^+_{ABC}\Gamma^{ABC}\psi_M^i\nonumber\\
		&-\frac{1}{4}\cancel{{D}}\phi^{ij}\psi_{Mj}+\phi^{ij}\phi_{Mj},\nonumber\\
		{D}_M\phi^{ij}&=(\partial_M-2b_M)\phi^{ij}+V_M{}^{[i}{}_k\phi^{j]k}+4\bigg(\bar{\psi}_M^{[i}\psi^{j]}-\text{trace}\bigg).\nonumber\\
		H_{MNP}&=3\partial_{[M}B_{NP]}+3\bar{\psi}_{[M}\Gamma_{NP]}\psi-\frac{3}{2}\bar{\psi}^i_{[M}\Gamma_N\psi^j_{P]}\phi_{ij}.\nonumber\\
	\end{align}
	
	We also give the Bianchi identity which will be useful later:\footnote{There is a minor sign mistake in the second term of this equation as it appears in \cite{Bergshoeff:1999db}. We have corrected the sign in equation (2.10).}
	\begin{align}\label{BI}
		&{D}_{[A}H_{BCD]}+\frac{3}{2}\bar{\psi}\Gamma_{[AB}R(Q)_{CD]}=0\nonumber\\
	\end{align}
	\section{Dimensional Reduction Scheme}
	One can relate supergravity theories in different dimensions via dimensional reduction. This technique is helpful when a direct construction of a supergravity theory in the lower dimension is not possible due to some reason. In our case of interest the direct construction of $\mathcal{N}=2$ conformal supergravity in five dimensions is not possible due to the non-existence of a rigid superconformal algebra which in turn does not allow us to use the standard method of constructing a conformal supergravity theory whose starting point is the gauging of a rigid superconformal algebra. Hence we resort to the dimensional reduction technique. Even when direct construction of supergravity theories  in a lower dimension is possible, relating them to a supergravity theory in a higher dimension via dimensional reduction is sometimes helpful in discovering new supersymmetric invariants. This was done for example in \cite{Banerjee:2011ts}, where the authors related 5d $\mathcal{N}=1$ conformal supergravity and 4d $\mathcal{N}=2$ conformal supergravity via dimensional reduction. This allowed them to discover new supersymmetric invariants and address some open questions concerning BPS black holes in five dimensions. There is however a subtle difference between the work of \cite{Banerjee:2011ts} and our current work. The authors of \cite{Banerjee:2011ts} deal with the dimensional reduction of half maximal conformal supergravity. Hence the $\mathcal{N}=1$ standard Weyl multiplet in five dimensions reduces to the $\mathcal{N}=2$ standard Weyl multiplet along with a vector multiplet in four dimensions. The Kaluza Klein scalar and vector resides in this vector multiplet. However, we are dealing with maximal conformal supergravity. Hence the standard Weyl multiplet in six dimensions reduces to a dilaton Weyl multiplet in five dimensions without any additional matter multiplet. The Kaluza Klein scalar and vector reside inside the dilaton Weyl multiplet instead of any other matter multiplet.  
	
In this section, we will give the details of the Kaluza-Klein dimensional reduction scheme from six dimensions to five dimensions. Let us decompose the six dimensional world index $M$ as $M=(\mu,z)$ where $\mu=0,1,2,3,4$ denotes the world index in the five dimensional spacetime and $z$ denotes the world index on the circle along which we are compactifying (see the figure below). Similarly  the six dimensional local Lorentz index $A$ is decomposed as $A=(a,\underline{5})$ where $a=\underline{0},\underline{1}, \underline{2}, \underline{3}, \underline{4}$ denotes the local Lorentz index in the five dimensional spacetime.  
	\begin{figure}[t]
		\centering
		\includegraphics[width=.6\textwidth]{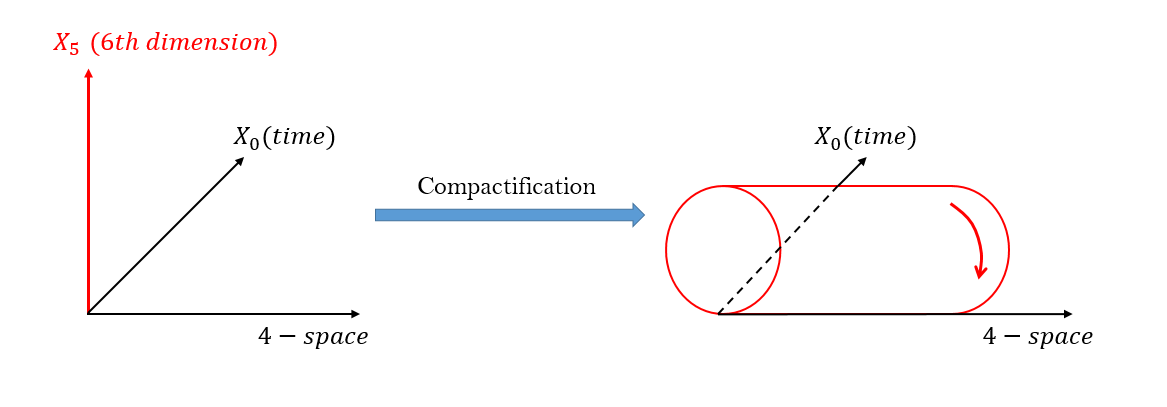}
		\caption{Six dimensions to five dimensions}
	\end{figure}
	We obtain an effective theory of massless fields in five dimensions by turning off the $z$-dependence of all the six dimensional fields. In order to obtain a theory of gravity in five dimensions we need to break the local Lorentz group of six dimensions down to five dimensions. This is obtained by setting $e_z^a=0$. This leads us to the following ansatz for the decomposition of the six dimensional vielbein to objects in five dimensions. 
	
	\begin{align}\label{vielbred}
		e^A_M= \begin{pmatrix}
			e^a_\mu & e^{\underline{5}}_\mu=\alpha^{-1}C_\mu\\
			e^a_z=0 & e^{\underline{5}}_z=\alpha^{-1}
		\end{pmatrix}\;,
	\end{align}
	and its inverse
	\begin{align}
		e_A^ M= \begin{pmatrix}
			e_a^\mu & e_a^z=-e^{\mu}_{a}C_\mu=-C_a\\
			e^\mu_{\underline{5}}=0 & e_{\underline{5}}^z=\alpha
		\end{pmatrix}\;.
	\end{align}

	The field $C_\mu$ appearing above is the Kaluza-Klein vector field which transforms as a gauge field under the local $U(1)$ that arises from the diffeomorphism along the internal circle in six dimensions. The field $\alpha$ is the Kaluza-Klein scalar. Since it transforms non-trivially under dilatation, it is also known as the dilaton scalar. Presence of this dilaton field makes the dimensionally reduced Weyl multiplet a dilaton Weyl multiplet.
	
	Similarly, we take $b_z=0$ in order to reduce the six dimensional special conformal transformation to five dimensions.
	
	It is to be noted that we did not set any constraint on the $\psi_z^i$ and $b_\mu$ fields as we do not wish to fix the S-supersymmetry and special conformal transformation as done in the paper \cite{Cordova:2013bea}. Because of this, the supergravity theory that we get is a theory of conformal supergravity whereas in \cite{Cordova:2013bea} it is a theory of Poincar\'e supergravity.

	We have already given the reduction ansatz of the six dimensional vielbein to five dimensional fields in (\ref{vielbred}). Below, we give the reduction ansatz of the remaining fields of the six dimensional Weyl multiplet to five dimensional fields. The fields appearing on the L.H.S are six dimensional fields whereas the fields appearing on the R.H.S are five dimensional fields.  
	\begin{align}\label{dic}
		\hat{b}_M&=\{b_\mu, 0\}\;,\nonumber \\
		\hat{V}_M^{ij}&=\left\{V_\mu^{ij} +  \alpha^{-1}C_\mu E^{ij}, \alpha^{-1}E^{ij}\right\}\;, \nonumber\\
		\hat{T}_{ab\underline{5}}^{ij}&=T_{ab}^{ij}\;, \nonumber\\
		\hat{T}_{abc}^{ij}&=-\frac{1}{2}\epsilon_{abcde}T^{deij}\;, \nonumber \\
		\hat{D}^{ij,kl}&=\frac{15}{8}D^{ij,kl}, \nonumber\\
		\hat{\psi}_M^i&=\left\{\ \begin{pmatrix}
			\psi_\mu^i + \alpha^{-1}C_\mu \lambda^i\\
			0
		\end{pmatrix}, \begin{pmatrix}
			\alpha^{-1}\lambda^i\\
			0
		\end{pmatrix}\right\}\;, \nonumber \\
		\hat{\chi}_k{}^{ij}&=\frac{15}{16}\begin{pmatrix}
			\chi_k{}^{ij}\\
			0
		\end{pmatrix} \nonumber\\
	\end{align}

	The six dimensional spinors appearing on the L.H.S above are 8 component symplectic Majorana left handed spinors which are decomposed into five dimensional spinors on the R.H.S which are 4-component symplectic Majorana spinors. The details of the dictionary between six dimensional and five dimensional spinors and Gamma matrices is given in Appendix-\ref{spin_Gamma_red}. In table \ref{table: 3}, the details of $\mathcal{N}=2$ dilaton Weyl multiplet in five dimensional conformal supergravity have been shown.

	\section{Dilaton Weyl Multiplet For $\mathcal{N}=2$ Conformal Supergravity In Five Dimensions}
 \begin{table}[t!]
		\centering
		\centering
		\begin{tabular}{|p{3cm}|p{7cm}|p{2.5cm}|p{2.5cm}|}
			\hline
			Field&Properties&USp(4) irreps& Weyl weight\\
			\hline
			\multicolumn{4}{|c|}{Independent Gauge fields}\\
			\hline
			$e_\mu^a$&vielbein&\bf{1}&-1\\
			$\psi_\mu^i$& Symplectic Majorana spinor&\bf{4}&-$\frac{1}{2}$\\
			$V_\mu^{ij}$&$V_\mu^{ij}=V_\mu^{ji}$&\bf{10}&0\\
			$C_\mu$& Graviphoton vector&\bf{1}&0\\
			\hline
			\multicolumn{4}{|c|}{Dependent Gauge fields}\\
			\hline
			$\omega_{\mu}{}^{ab}$& $\omega_{\mu}{}^{ab}=-\omega_{\mu}{}^{ba}$, Spin Connection&\bf{1}&0\\
			$\phi_{\mu}^i$&Symplectic Majorana Spinor &\bf{4}&$\frac{1}{2}$\\
			$f_\mu^a$&Special Conformal Transformation Gauge Field&\bf{1}&1\\
			\hline
			\multicolumn{4}{|c|}{Auxiliary matter fields}\\
			\hline
			$\alpha$& Real Scalar&\bf{1}&1\\
			$E^{ij}$ & $E^{ij}=E^{ji}$ & \bf{10} & 1\\
			$\lambda^i$&Symplectic Majorana spinor&\bf{4}&$\frac{1}{2}$\\
			$T^{ij}_{ab}$  & $T^{ij}_{ab}=-T^{ji}_{ab}$, $T^{ij}_{ab}=-T^{ij}_{ba}$, $\Omega_{ij}T^{ij}_{ab}=0$&\bf{5}&1\\
			$D^{ij,kl}$& $D^{ij,kl}=-D^{ji,kl}=-D^{ij,lk},D^{ij,kl}=D^{kl,ij}, \Omega_{ij}D^{ij,kl}=0,{\Omega_{ik}\Omega_{jl}D^{ij,kl}=0}$&\bf{14}&2\\
			$\chi_k{}^{ij}$ &Symplectic Majorana, $\chi_k{}^{ij}=-\chi_k{}^{ji}, \Omega_{ij}\chi_k{}^{ij}=0, \chi_i{}^{ij}=0$&\bf{16}&$\frac{3}{2}$\\
			\hline
		\end{tabular}
		\caption{Five Dimensional Dilaton Weyl Multiplet}
		\label{table: 3}	
	\end{table}
	
	In this section, we will obtain all the gauge transformations of the dilaton Weyl multiplet from the six dimensional conformal supergravity by using the dimensional reduction scheme. Before going into details of the transformations, it is to be noted that the gauge fixing conditions $e_{z}^{a}=0=b_z$ is not preserved by the six dimensional supersymmetry transformations (both Q as well as S). Hence, the six dimensional supersymmetry transformations will not directly reduce to the five dimensional supersymmetry transformations. Instead what will reduce to the five dimensional supersymmetry transformations is a combination of the six dimensional supersymmetry transformation and a field dependent local Lorentz and special conformal transformations given by the parameters $\Lambda^{a{\underline{5}}}$ and $\Lambda^{\underline{5}}$.
	Hence, the following combination of six dimensional supersymmetry and field dependent gauge transformations is what will reduce to the correct five dimensional supersymmetry transformations: 
	\begin{align}\label{6d-5d}
		[\delta]_{\text{6D}}+[\delta_{L}]_{\text{6D}}(\Lambda^{a{\underline{5}}}) + [\delta_{\text{SCT}}]_{6D}(\Lambda_{\underline{5}})\to [\delta]_{\text{5D}}\;,
	\end{align} 
	where $\delta$ refers to both Q as well as S-supersymmetry transformation and the field dependent parameters appearing above are given as follows:
	{\allowdisplaybreaks
		\begin{align}
			\Lambda^{a{\underline{5}}}&=-\frac{1}{2}\bar{\epsilon}\gamma^a\lambda\;, \nonumber \\
			\Lambda_{\underline{5}}&=\frac{i\alpha}{4}\bar{\epsilon}\phi_z+\frac{i}{4\alpha}\bar{\eta}\lambda\;.
		\end{align}
	}
	
	The gauge transformations of the independent fields of the Weyl multiplet in five dimension can be obtained from the 6d-5d dictionary of the components of the Weyl multiplet as given in (\ref{dic}) and the 6d-5d dictionary of the supersymmetry transformations as given in (\ref{6d-5d}). 
	
	Below we note down the gauge transformations of the independent gauge fields, where $\epsilon$, $\eta$, $\Lambda_{U}$ $\Lambda_D$, $\Lambda_a$, $\Lambda^{a}{}_{b}$, $\Lambda^{ij}$ denotes the parameters of Q-supersymmetry, S-supersymmetry, $U(1)$, dilatation, special conformal transformation, local Lorentz transfornation and $USp(4)$ R-symmetry respectively:
	{\allowdisplaybreaks
		\begin{align}\label{5dsusy}
			\delta e_{\mu}{}^a &= -\Lambda_D e_\mu^a -\Lambda^{ab}e_{\mu b} -\frac{1}{2}\bar{\epsilon}\gamma^a \psi_\mu\nonumber\\
			\delta \psi^i_{\mu} &=-\frac{1}{2}\Lambda_D \psi_\mu^i +\frac{1}{2}\Lambda^i_{~j}\psi^j_\mu -\frac{1}{4}\Lambda^{ab}\gamma_{ab}\psi_\mu^i+ \mathcal{D}_\mu \epsilon^i 
			+ \frac{i}{4\alpha}G_\mu{}^{a}\gamma _a \epsilon ^i 
			-\frac{i}{4}\bar{\psi}_\mu \gamma^a \lambda \gamma_a \epsilon^i
			\nonumber\\
			&-\frac{i}{4}\gamma\cdot T^{~ij}  \gamma_\mu \epsilon_j
			+\frac{i}{2} \bar{\epsilon}\psi_\mu\lambda^i+\frac{1}{2}\bar{\epsilon}\gamma_\mu\lambda \lambda^i 
			+\frac{i}{4}\bar{\epsilon}\gamma ^a \lambda \gamma_a \psi_\mu ^{~i}+i\gamma_\mu\eta^i\nonumber\\  
			\delta b_\mu &= \partial_\mu \Lambda_D -2e^a_\mu \Lambda_{a}+\frac{i}{2}\Bar{\epsilon}\phi_\mu+\frac{i}{2}\bar{\eta}\psi_\mu\nonumber\\
			\delta V_\mu ^{ij} &= \partial_\mu \Lambda^{ij}+\Lambda ^{(i}{}_{k}V_\mu ^{~j)k} +4i\Bar{\epsilon}^{(i}\phi_\mu^{j)} 
			+\frac{1}{4}\bar{\epsilon}_k \gamma_\mu \chi^{(i,j)k} 
			+\frac{1}{2} E^{ij}(i\bar{\epsilon}\psi_\mu 
			+ \bar{\epsilon}\gamma_\mu \lambda)-4i\bar{\eta}^{(i}\psi_\mu^{j)}\nonumber\\
			\delta C_\mu &= \partial_\mu \Lambda_U-\frac{i}{2}\alpha \bar{\epsilon}\psi_\mu 
			-\frac{\alpha}{2} \bar{\epsilon}\gamma_\mu \lambda\;,
		\end{align}	
	}

	where
	\begin{align}
		\mathcal{D}_\mu \epsilon^i &= \partial_\mu \epsilon^i +\frac{1}{2}b_\mu \epsilon^i +\frac{1}{4}\gamma\cdot\omega_\mu  \epsilon^i- \frac{1}{2}V_\mu{}^{i}_{~j}\epsilon^j\;,
	\end{align}
	$G_{\mu\nu}$ denotes the fully supercovariant curvature corresponding to the gauge field $C_{\mu}$ given in (\ref{SupCovCurv}) and $D_{\mu}$ denotes the fully supercovariant derivative. We have also used the notation $\chi^{i,jk}$ which is defined as follows: $\chi^{i,jk}=\Omega^{il}\chi_l{}^{jk}$.		
	
	Below we note down the gauge transformations of the auxiliary matter fields ($\alpha$, $\lambda^{i}$, $E^{ij}$, $T_{ab}^{ij}$):	
	{\allowdisplaybreaks
		\begin{align}
			\delta \alpha &= \Lambda_D \alpha+\frac{i}{2}\alpha \bar{\epsilon}\lambda\nonumber\\
			\delta \lambda^i &= \frac{1}{2}\Lambda_D \lambda^i+\frac{1}{2}\Lambda^i{}_j\lambda^j-\frac{1}{4}\Lambda^{ab}\gamma_{ab}\lambda^j-\frac{1}{8\alpha}\gamma \cdot G \epsilon^i 
			+ \frac{i}{2\alpha} \cancel{D} \alpha \epsilon^i 
			-\frac{1}{2}E^i_{~j}\epsilon^j+\frac{1}{4}\gamma\cdot T^{ij}\epsilon_j\nonumber\\
			&+\frac{i}{2}\bar{\epsilon}\lambda\lambda^i 
			+\frac{i}{4}\bar{\epsilon}\gamma ^a \lambda \gamma _a \lambda^i+\eta^i\nonumber\\ 
			\delta E^{ij} &=\Lambda_D E^{ij}+\Lambda^{(i}{}_kE^{j)k}+\frac{4}{5\alpha}\bar{\epsilon}^{(i} \lambda^{j)} \cancel{D}\alpha 
			-\frac{4}{5}\epsilon^{(i}\cancel{D}\lambda^{j)}-\frac{i}{5\alpha}\epsilon^{(i}\gamma\cdot G \lambda^{j)}
			+\frac{i}{5}\bar{\epsilon}^{(i}\gamma \cdot T^{j)k} \lambda_k
			\nonumber \\
			&+\frac{i}{4}\bar{\epsilon}_k \chi^{(i,j)k}+\frac{i}{2}E^{ij}\bar{\epsilon}\lambda-4i\bar{\eta}^{(i}\lambda^{j)}\nonumber\\
			\delta T^{ij}_{ab}&=\Lambda T_{ab}^{ij}+2T^{ij}_{c[a}\Lambda^c{}_{b]}-\Lambda^{[i}{}_kT^{j]k}_{ab}-\frac{i}{8}\bar{\epsilon}^{[i}\gamma^{cd}\gamma_{ab}R(Q)_{cd}^{j]}+\frac{i}{16}\bar{\epsilon}^k\gamma_{ab}\chi_k{}^{ij}\nonumber\\
		&-\frac{i}{8\alpha}\bar{\epsilon}^{[i}\gamma \cdot G\gamma_{ab}\lambda^{j]}-\frac{i}{4}\bar{\epsilon}\gamma_{abcd}\lambda T^{cd,ij}-\frac{1}{4}\bar{\epsilon}^{[i}\gamma^c\gamma_{ab}\bigg(-\frac{1}{\alpha}D_c\alpha\lambda^{j]}+D_c\lambda^{j]}\nonumber\\
			&+\frac{i}{4\alpha}G_{cd}\gamma^d\lambda^{j]}-\frac{i}{4}\gamma\cdot T^{j]k} \gamma_c\lambda_k\bigg)-\frac{1}{20}\bar{\epsilon}^{[i}\gamma_{ab}\bigg(\frac{1}{\alpha}\cancel{D}\alpha\lambda^{j]}-\cancel{D}\lambda^{j]}-\frac{i}{4\alpha}\gamma\cdot G \lambda^{j]}\nonumber\\
			&+\frac{i}{4}\gamma\cdot T^{j]k}\lambda_k\bigg)-\Omega\text{trace}\nonumber\\
		\end{align}
	}

	where $R(Q)_{\mu\nu}^{i}$ denotes the supercovariant curvature corresponding to the gauge field $\psi_{\mu}^{i}$ of Q-supersymmetry (\ref{SupCovCurv}). Below we note down the gauge transformations of the auxiliary matter fields ($\chi_{k}{}^{ij}$ and $D^{ij;kl}$):
	{\allowdisplaybreaks
		\begin{align}\label{chi_D}
			\delta \chi_{k}{}^{ij}&=\frac{3}{2}\Lambda_D\chi_{k}{}^{ij}-\frac{1}{4}\Lambda^{ab}\gamma_{ab}\chi_k{}^{ij}-\Lambda^{[i}{}_l\chi_k{}^{j]l}-\frac{1}{2}\chi_l{}^{ij}\Lambda^l{}_k-iD_c T_{ab}^{ij}\gamma^{ab}\gamma^c\epsilon_k-\gamma \cdot R(V)_{k}{}^{[i}\epsilon^{j]}\nonumber\\
			&-\frac{1}{2}D_{kl}^{ij}\epsilon^l+4\gamma \cdot T^{ij}\eta_k+\frac{1}{2\alpha}\gamma_{abc}T^{ab,ij}\bigg(G^{cd}\gamma_d+2iD^c\alpha\bigg)\epsilon_k+ \frac{1}{\alpha}{G}_{a}{}^{d}T_{bd}^{ij}\gamma^{ab}\epsilon_k\nonumber \\
			&+\frac{i}{8}\bar{\lambda}^{[i}\gamma^{cd}\gamma_{ab}R(Q)_{cd}^{j]}\gamma^{ab}\epsilon_k+\frac{11i}{20\alpha}\bar{\lambda}^{[i}\gamma_{b}{}^{d} G_{ad}\lambda^{j]}\gamma^{ab}\epsilon_k-\frac{1}{5\alpha}\bar{\lambda}^{[i}\gamma_{ab}\cancel{D}\alpha\lambda^{j]}\gamma^{ab}\epsilon_k\nonumber\\
			&+\frac{1}{4}\bar{\lambda}^{[i}\gamma^c\gamma_{ab}\bigg(D_c\lambda^{j]}-\frac{i}{4}\gamma\cdot T^{j]l} \gamma_c\lambda_l\bigg)\gamma^{ab}\epsilon_k+\frac{1}{20}\bar{\lambda}^{[i}\gamma_{ab}\bigg(-\cancel{D}\lambda^{j]}+\frac{i}{4}\gamma\cdot T^{j]l}\lambda_l\bigg)\gamma^{ab}\epsilon_k\nonumber\\
			&+E^{[i}{}_{l}\gamma\cdot T^{j]l}\epsilon_k-\frac{i}{16}\bar{\lambda}^m\gamma_{ab}\chi_m{}^{ij}\gamma^{ab}\epsilon_k-\frac{1}{\alpha}\gamma \cdot{G}E_{k}{}^{[i}\epsilon^{j]}-2i\cancel{D}E_{k}{}^{[i}\epsilon^{j]}\nonumber\\
			&+\frac{2i}{\alpha}E_{k}{}^{[i}\cancel{D}\alpha \epsilon^{j]}+\frac{i}{2}\bar{\lambda}_m\gamma_a\chi_{(k,l)}{}^{m}\gamma^a \Omega^{l[i}\epsilon^{j]}+\frac{i}{4}\bar{\epsilon}\gamma^a\lambda\gamma_a\chi_k{}^{ij}-\text{traces}\nonumber\\
			\delta {D^{ij,kl}}&=\Lambda_D D^{ij,kl}-\Lambda^{[i}{}_{m}D^{j]m,kl}+\bar{\epsilon}^{[i}\cancel{D}\chi^{j],kl}-\frac{3i}{8\alpha}\bar{\epsilon}^{[i}\gamma\cdot G\chi^{j],kl}-\frac{i}{2\alpha}G_{cd}\bar{\epsilon}^{[i}\gamma^{abc}\gamma^d \lambda^{j]} T_{ab}^{kl}\nonumber\\
			&+i\bar{\epsilon}^{[i}\bigg(\chi^{j],m[k}E_m{}^{l]}+\frac{1}{2}E^{j]m}\chi_m{}^{kl}-\frac{1}{2}D^{j]m,kl}\lambda_m\bigg)-\frac{1}{2\alpha}\bar{\epsilon}^{[i}\cancel{D}\alpha\chi^{j],kl}-\bar{\epsilon}^{[i}\gamma^{ab} \gamma^c\lambda^{j]}{D}_cT_{ab}^{kl} \nonumber\\
			&+\frac{1}{\alpha}D_a\alpha\bar{\epsilon}^{[i}\gamma^{abc} \lambda^{j]} T_{bc}^{kl}-i\bar{\epsilon}^{[i} \gamma^{ab}\lambda^{j]}\bigg \{\frac{1}{\alpha}{G}_{a}{}^{d}T_{bd}^{kl}-\frac{1}{5\alpha}\bar{\lambda}^{[k}\gamma_{ab}\cancel{D}\alpha\lambda^{l]}\nonumber\\
			&{+\frac{i}{8}\bar{\lambda}^{[k}\gamma^{cd}\gamma_{ab}R(Q)_{cd}^{l]}+\frac{11i}{20\alpha}\bar{\lambda}^{[k}\gamma_{bc} G_{a}{}^c\lambda^{l]}} {+\frac{1}{4}\bar{\lambda}^{[k}\gamma^c\gamma_{ab}\bigg(D_c\lambda^{l]}-\frac{i}{4}\gamma \cdot T^{l]n}\gamma_c\lambda_n\bigg)}\nonumber\\
			&{+\frac{1}{20}\bar{\lambda}^{[k}\gamma_{ab}\bigg(-\cancel{D}\lambda^{l]}+\frac{i}{4}\gamma\cdot T^{l]n}\lambda_n\bigg)}+E^{[k}{}_{m}T^{l]m}_{ab}-\frac{i}{16}\bar{\lambda}^m\gamma_{ab}\chi_m^{kl}-\Omega\text{trace}\bigg\}\nonumber\\
			&-i\bar{\epsilon}^{[i}\Omega^{j]m}\bigg\{-\gamma \cdot R_{\mu \nu, m}{}^{[k}\lambda^{l]}-\frac{1}{\alpha}\gamma\cdot{G}E_{m}{}^{[k}\lambda^{l]}-2i\cancel{D} E_{m}{}^{[k}{}\lambda^{l]}+\frac{2i}{\alpha}E_{m}{}^{[k}\cancel{D}\alpha \lambda^{l]}\nonumber\\
			&-\frac{1}{2}\bar{\lambda}_r\gamma_c\chi_{(m,q)}{}^{r}\gamma^c\Omega^{q[k}\lambda^{l]}-\frac{4i}{5}\gamma \cdot T^{kl}\bigg(\frac{1}{\alpha}\cancel{D}\alpha \lambda_m-\cancel{D}\lambda_m-\frac{i}{4\alpha}\gamma\cdot G\lambda_m\nonumber\\
			&-\frac{i}{4}\gamma\cdot T_{mn}\lambda^{n}\bigg)\bigg\}+2i\bar{\eta}^{[i}\chi^{j],kl}-\text{traces}\nonumber\\
		\end{align}
	}
	
	where $R(V)_{\mu\nu}{}^{ij}$ denotes the fully supercovariant curvature associated with the $USp(4)$ gauge field $V_{\mu}{}^{ij}$ as given in \ref{SupCovCurv} where we note down the expressions for all the fully super-covariant curvatures.

	The expressions for the fully super-covariant curvatures are as follows:
	{\allowdisplaybreaks
		\begin{align}\label{SupCovCurv}
			G_{\mu \nu} &= 2\partial_{[\mu} C_{\nu]}+\frac{i\alpha}{2}\Bar{\psi}_\mu \psi_\nu+\alpha\Bar{\psi}_{[\mu} \gamma_{\nu]} \lambda \nonumber \\
			R(P)_{\mu \nu}{}^a &= 2\partial_{[\mu}e_{\nu]}^a+2b_{[\mu}e_{\nu]}^a+2\omega_{[\mu}^{ab}e_{\nu]b}+\frac{1}{2}\bar{\psi}_\mu \gamma^a \psi_\nu \nonumber 
			\\
			R(Q)_{\mu\nu}{}^{i}&=2\partial_{[\mu}\psi^i_{\nu]}+b_{[\mu}\psi^i_{\nu]}+\frac{1}{2}\omega_{[\mu}^{ab}\gamma_{ab}\psi_{\nu]}^i+V_{[\mu}^{ij} \psi_{\nu]j} +\frac{i}{2\alpha}G_{[\mu}^{~~a}\gamma_a \psi_{\nu]}^i -\frac{i}{2}\Bar{\psi}_{[\mu}\gamma^a\lambda\gamma_a\psi_{\nu]}^i\nonumber  \\
			& -\frac{i}{2}T^{ij}_{ab}\gamma^{ab}\gamma_{[\mu}\psi_{\nu]j}
			-\frac{i}{2}\Bar{\psi}_{[\mu}\psi_{\nu]}\lambda^i -\Bar{\psi}_{[\mu}\gamma_{\nu]} \lambda\lambda^i+2i\gamma_{[\mu}\phi_{\nu]}^i \nonumber \\
			R(M)_{\mu \nu }{}^{ab}&=2\partial_{[\mu}\omega_{\nu]}{}^{ab}+2\omega_{[\mu}{}^{ac}\omega_{\nu]c}{}^b+8e_{[\nu}{}^{[a}f_{\mu]}{}^{b]} -\frac{i}{4}\Bar{\psi}_{[\mu}^i \gamma^{[a}\gamma_{cd}\gamma^{b]}\psi_{\nu]}^jT^{cd}_{ij}+\frac{i}{4\alpha}\Bar{\psi}_{[\mu} \psi_{\nu]} G^{ab}\nonumber\\
			&+\frac{1}{8}\bigg(\Bar{\lambda}\gamma^{ab}\lambda\Bar{\psi}_{[\mu}\psi_{\nu]}+\Bar{\lambda}\gamma^{[a}\gamma_c\gamma^{b]} \lambda \Bar{\psi}_{[\mu}\gamma^c\psi_{\nu]}\bigg)-i\Bar{\psi}_{[\mu}\gamma^{ab}\phi_{\nu]} -\frac{1}{2}\bar{\psi}_{[\mu}\gamma_{\nu]} R(Q)^{ab}\nonumber\\
			&-\bar{\psi}_{[\mu}\gamma^{[a}R(Q)_{\nu]}{}^{b]}\nonumber\\
			R(V)_{\mu \nu}{}^{ij}&=2\partial_{[\mu}V_{\nu ]}{}^{ij}+{V_{[\mu}{}^{k(i}V_{\nu]}{}^{j)}{}_k}-8i\Bar{\epsilon}^{(i}\phi_\mu^{j)} -\frac{1}{2}\bar{\psi}_{[\mu, k} \gamma_{\nu]} \chi^{(i,j)k} 
			-E^{ij}\bigg(\frac{i}{2}\bar{\psi}_{[\mu}\psi_{\nu]} 
			+ \bar{\psi}_{[\mu}\gamma_{\nu]} \lambda\bigg)\nonumber\\
	\end{align} }
	
	The  gauge fields $\omega_\mu^{ab}, f_\mu^a, \phi_\mu^i$ are dependent due to the curvature constraints as shown below. 
	{\allowdisplaybreaks
		\begin{align}\label{curv_const1}
			R(P)_{\mu \nu}^a &= 0 \nonumber
			\\
			R(M)_{\mu\nu}^{~~~ab}e^\nu_{~~b}+\frac{1}{2}T_{\mu b}^{~~ij}T_{ij}^{~~ab} &= 0\nonumber \\
			\gamma^a R(Q)_{ab}^{i}&=\zeta^i_b
			\,.
		\end{align}
		
		where 
		\begin{align}
			\zeta^i_a&=-\frac{i}{\alpha}\bigg(D_a\alpha-\frac{1}{5}\gamma_a\cancel{D}\alpha\bigg)\lambda^i+i\bigg(D_a\lambda^i-\frac{1}{5}\gamma_a\cancel{D}\lambda^i\bigg)\nonumber\\
			&+\frac{1}{4\alpha}\bigg(3G_{ab}\gamma^b+\frac{1}{5}\gamma_a \gamma\cdot G\bigg)\lambda^i-\frac{1}{4}\bigg(\gamma\cdot T^{ij}\gamma_a-\frac{1}{5}\gamma_aT^{ij}\cdot\gamma\bigg)\lambda_j\nonumber\\
		\end{align}
		The expressions for the dependent fields can be obtained by solving the curvature constraints given in (\ref{curv_const1}):
		{\allowdisplaybreaks
			\begin{align}
				\omega_\mu{}^{ab} &=2e^{\nu[a}\partial_{[\mu}e_{\nu]}^{~b]}-e^{\rho[a}e^{b]\sigma}e_\mu^{~c}\partial_\rho e_{\sigma c}+2e_\mu^{[a}b^{b]}-\frac{1}{2}\Bar{\psi}_\mu \gamma^{[a}\psi^{b]}-\frac{1}{4}\Bar{\psi}^a \gamma_\mu \psi^b\nonumber \\
				\phi^i _\mu &= \frac{i}{12}\bigg(\gamma^{ab}\gamma_\mu -\frac{1}{2}\gamma_\mu \gamma^{ab}\bigg) R^{'}{}_{ab}{}^i+\frac{i}{24}\gamma_\mu \gamma\cdot G\lambda^i-\frac{i}{3}\zeta_\mu^i \nonumber\\
				f_\mu ^{~a} &= \frac{1}{48}\bigg(8R^{'a}_\mu -R'e_\mu{}^a-4T_{\mu b}^{~~ij}T_{ij}^{~~ab}-\frac{1}{2}e_\mu^a T^{ij}_{ab}T^{ab}_{ij}\bigg)\nonumber\\
			\end{align}
			In the above expressions, we have defined the following:
			{\allowdisplaybreaks
				\begin{align}
					R^{'}(M)_{\mu \nu}{}^{ab}&=R(M)_{\mu\nu}{}^{ab}|_{f=0}\nonumber \\
					R^{'}(Q)_{\mu\nu}^{i}&=R(Q)_{\mu\nu}^{i}|_{\phi=0} \nonumber \\
					R^{'}(M)_{\mu}^{a}&=e^{\nu}_{b}R^{'}(M)_{\mu \nu}{}^{ab}\;, R^{'}(M)=e^{\mu}_{a}R^{'}(M)_{\mu}^{a}
			\end{align}}

			The soft-algebra realized on the Weyl multiplet is as follows:
			{\allowdisplaybreaks
				\begin{align}
					[\delta_{Q_1} , \delta_{Q_2}] &= \delta_Q\bigg(\epsilon_3^{(Q,Q)}{}^i\bigg)+\delta_{U(1)}\bigg(\Lambda_{3U}^{(Q,Q)}\bigg)+ \delta_{c.g.c.t}\bigg(\zeta_3^{(Q,Q)}{}^a\bigg) +\delta_{USp}\bigg(\Lambda_3^{(Q,Q)}{}^{ij}\bigg)\nonumber\\
					&+\delta_S\bigg(\eta_3^{(Q,Q)}{}^i\bigg)+ \delta_{S.C.T}\bigg(\Lambda^{(Q,Q)}_{3,a}\bigg)+\delta_L\bigg(\Lambda_{3}^{(Q,Q)}{}^{ab}\bigg) \nonumber\\
					[\delta_{S_1} , \delta_{S_2}] &= \delta_{S.C.T}\bigg(\Lambda^{(S,S)}_{a}\bigg) \nonumber\\
					[\delta_{Q} , \delta_{S}] &=\delta_S\bigg(\eta^{(Q,S)}_3{}^i\bigg)+ \delta_{USp}\bigg(\Lambda_3^{(Q,S)}{}^{ij}\bigg) 
					+\delta_L\bigg(\Lambda_3^{(Q,S)}{}^{ab}\bigg)+ \delta_{K}\bigg(\Lambda^{(Q,S)}_{D3}\bigg)+\delta_{S.C.T}\bigg(\Lambda_{3}^{(Q,S)}{}^a\bigg)\nonumber\\
			\end{align}}

			The field dependent parameters in the above soft algbera are given below:
			{\allowdisplaybreaks
				\begin{align}
					\epsilon_3^{(Q,Q)}{}^i &= -\frac{i}{2}\Bar{\epsilon}_1 \epsilon_2 \lambda^i-\frac{i}{4}\Bar{\epsilon}_1\gamma_a \lambda \gamma^a \epsilon_2^i+\frac{i}{4}\Bar{\epsilon}_2\gamma^a\lambda\gamma_a\epsilon_1^i\nonumber\\
					\Lambda^{(Q,Q)}_{3U}&=-\frac{i}{2}\alpha\bar{\epsilon}_2\epsilon_{1}\nonumber\\
					\zeta_3^{(Q,Q)}{}^a &=\frac{1}{2}\Bar{\epsilon_1}\gamma^a \epsilon_2\nonumber\\
					\Lambda_3^{(Q,Q)}{}^{ij} &= -\frac{i}{2}E^{ij}\Bar{\epsilon}_1\epsilon_2\nonumber\\
					\eta_3^{(Q,Q)}{}^i 
					&=\biggl[-\gamma^{ab}\epsilon_2^i\bigg(-\frac{i}{16}\bar{\epsilon}_1R(Q)_{ab}+\frac{1}{8}\bar{\epsilon}_1\gamma_a D_b\lambda-\frac{1}{8\alpha}D_a\alpha \bar{\epsilon}_1\gamma_b\lambda+\frac{i}{32\alpha}\bar{\lambda}\gamma_a\gamma_c G_{bc}\epsilon_1\nonumber\\
					&+\frac{i}{32}\bar{\lambda}^j\gamma_a\gamma\cdot T_{jk}\gamma_b\epsilon_{1}^k\bigg)-\gamma^a\epsilon_2^i\bigg(\frac{1}{4}\bar{\epsilon}_1D_a\lambda+\frac{i}{16\alpha}\bar{\lambda}G_{ac}\gamma^c\epsilon_{1}+\frac{i}{16}\bar{\lambda}^j\gamma\cdot T_{jk}\gamma_a\epsilon_{1}^k\bigg)\nonumber\\
					&-\frac{1}{4}\gamma^{ab}\epsilon_{2j}\biggl\{\frac{i}{8\alpha}\bar{\epsilon}_1^{[i}\gamma \cdot G\gamma_{ab}\lambda^{j]}+\frac{1}{4}\bar{\epsilon}_1^{[i}\gamma^c\gamma_{ab}\bigg(-\frac{1}{\alpha}D_c\alpha\lambda^{j]}+D_c\lambda^{j]}+\frac{i}{4\alpha}G_{cd}\gamma^d\lambda^{j]}\nonumber\\
					&-\frac{i}{4}T^{j]k}\cdot\gamma \gamma_c\lambda_k\bigg)+\frac{1}{20}\bar{\epsilon}_1^{[i}\gamma_{ab}\bigg(\frac{1}{\alpha}\cancel{D}\alpha\lambda^{j]}-\cancel{D}\lambda^{j]}-\frac{i}{4\alpha}G\cdot\gamma \lambda^{j]}+\frac{i}{4}T^{j]k}\cdot\gamma\lambda_k\bigg)\nonumber\\
					&-\frac{i}{16}\bar{\epsilon}_1^k\gamma_{ab}\chi_k^{ij}+\frac{i}{4}\bar{\epsilon}_1\gamma_{abcd}\lambda T^{cd,ij}-\frac{i}{8}\bar{\epsilon}^{[i}\gamma^{cd}\gamma_{ab}R(Q)_{cd}^{j]}-\Omega\text{trace}\biggr\}
					\nonumber\\
					&+\frac{i}{4}\bar{\epsilon}_2\gamma^a\lambda\gamma_a\bigg(-\frac{1}{8\alpha} G \cdot \gamma \epsilon_1^i + \frac{i}{2\alpha} D^a \alpha \cdot \gamma_a \epsilon_1^i-\frac{1}{2}E^i_{~j}\epsilon_1^j 
					+\frac{1}{4}\gamma\cdot T^{ij} \epsilon_{1,j} 
						\nonumber\\
					&+ \frac{i}{4}\bar{\epsilon}_1\gamma ^a \lambda \gamma _a \lambda^i+\frac{i}{2}\bar{\epsilon}_1\lambda\lambda^i\bigg)
					-\frac{1}{2}\epsilon_{2j}\bigg(-\frac{4}{5\alpha}\bar{\epsilon}_1^{(i} \lambda^{j)} \cancel{D}\alpha +\frac{4}{5}\bar{\epsilon}^{(i}_1\cancel{D}\lambda^{j)}-i\bar{\epsilon}_1^{(i}\gamma\cdot T^{j)k}  \lambda_k
				\nonumber\\
					&	-\frac{i}{4}\bar{\epsilon}_{1,k}\chi^{(i,j)k}
					-\frac{i}{2}E^{ij}\bar{\epsilon}_1\lambda\bigg)+\frac{i}{2}\bar{\epsilon}_2\lambda\bigg(-\frac{1}{8\alpha}\gamma \cdot G \epsilon_1^i 
					+ \frac{i}{2\alpha} \cancel{D} \alpha \epsilon_1^i 
					-\frac{1}{2}E^i_{~j}\epsilon_1^j
				\nonumber\\
					&+\frac{1}{4}\gamma\cdot T^{ij}\epsilon_{1,j}+\frac{i}{2}\bar{\epsilon}_1\lambda\lambda^i 	+\frac{i}{2}\bar{\epsilon}_1\gamma ^a \lambda \gamma _a \lambda^i\bigg)+\gamma_a\lambda^i\bigg(-\frac{1}{16}\bar{\epsilon}_1\gamma_b\lambda\bar{\epsilon}_2\gamma^a\gamma^b\lambda\nonumber\\
					&-\frac{1}{8}\bar{\epsilon}_1\lambda\bar{\epsilon}_2\gamma^a\lambda\bigg)-(1\leftrightarrow2)\biggr]+i\lambda^i\bigg(\frac{i}{2\alpha}\bar{\epsilon}_2\cancel{D}\alpha\epsilon_1-\frac{1}{4}\bar{\epsilon}_2^j\gamma\cdot T_{jk}\epsilon^k_1\bigg)\nonumber\\
					&-\gamma_b\lambda^i\bigg(\frac{i}{8\alpha}\bar{\epsilon}_2\gamma_a\epsilon_{1}G^{ab}+\frac{1}{4\alpha}\bar{\epsilon}_2D^b\alpha\epsilon_{1}+\frac{i}{4}\bar{\epsilon}_2^j\gamma_a\epsilon_1^k T^{ab}_{jk}\bigg)-\frac{i}{32\alpha}\bigg(G\cdot\gamma-4i\cancel{D}\alpha\bigg)\nonumber\\
					&\bigg(2\Bar{\epsilon}_1 \epsilon_2 \lambda^i+\Bar{\epsilon}_1\gamma_a \lambda \gamma^a \epsilon_2^i- \Bar{\epsilon}_2\gamma^a\lambda\gamma_a\epsilon_1^i\bigg)+\frac{i}{16}\bigg(2E^{ij}+T^{ij}\cdot\gamma\bigg)\bigg(2\Bar{\epsilon}_1 \epsilon_2 \lambda_j\nonumber\\
					&+\Bar{\epsilon}_1\gamma_a \lambda \gamma^a \epsilon_{2,j}- \Bar{\epsilon}_2\gamma^a\lambda\gamma_a\epsilon_{1,j}\bigg)-\frac{1}{16}\bigg(\bar{\epsilon}_1\gamma^b\lambda \bar{\lambda}\gamma_a\gamma_b\epsilon_2-\bar{\epsilon}_2\gamma^b\lambda \bar{\lambda}\gamma_a\gamma_b\epsilon_1\bigg)\gamma^a\lambda^i\nonumber\\
					&-\frac{i}{32\alpha}\bigg\{-2\alpha\Bar{\epsilon}_2^i \gamma^{[a}\gamma_{cd}\gamma^{b]}\epsilon_1^jT^{cd}_{ij}+2\Bar{\epsilon}_2 \epsilon_1 G^{ab}-i\alpha\bigg(\Bar{\lambda}\gamma^{ab}\lambda\Bar{\epsilon}_2\epsilon_1\nonumber\\
					&+\Bar{\lambda}\gamma^{[a}\gamma_c\gamma^{b]} \lambda \Bar{\epsilon}_2\gamma^c\epsilon_1\bigg)\bigg\}\gamma_{ab}\lambda^i-\frac{i}{4}E^{ij}\lambda_j\bar{\epsilon}_1\epsilon_2-\frac{1}{2}\bar{\epsilon}_1\gamma^a\epsilon_2D_a\lambda\nonumber\\
					\Lambda_{3}^{(Q,Q)}{}^a&=-\frac{1}{24} T^{ac}_{ij}T^{ij}_{bc}\bar{\epsilon}_2\gamma^b\epsilon_{1}+\frac{1}{384}\bar{\epsilon}_2\bigg(\gamma^a\gamma_{cd}\gamma^{ef}-\gamma^{ef}\gamma_{cd}\gamma^a\bigg)\epsilon_{1}R(M)_{cd,ef}\nonumber\\
					&-\frac{i}{48}\bar{\epsilon}_1\gamma_b\epsilon_2\bar{\lambda}R(Q)^{ab}-\frac{1}{48}\bar{\epsilon}_1\gamma_b\epsilon_2\bar{\lambda}^i\bigg\{\gamma^{[a}\bigg(\frac{1}{2}D^{b]}\lambda_i-\frac{3i}{8\alpha}G^{b]c}\gamma_c \lambda_i\nonumber\\
					&+\frac{i}{8}T_{ij}\cdot\gamma \gamma^{b]}\lambda_{j}\bigg)+\gamma^{ab}\bigg(\frac{1}{10\alpha}\cancel{D}\alpha\lambda_i-\frac{1}{10}\cancel{D}\lambda^i+\frac{i}{40\alpha}G\cdot \gamma\lambda^i-\frac{i}{40}T_{ij}\cdot\gamma\lambda^{j}\bigg)\bigg\}\nonumber\\
					&+\bigg[-\frac{i}{4}\bar{\epsilon}_2^i\bigg\{-\frac{i}{3}\gamma_b\bigg(\frac{i}{48}\gamma_c G^{c[a}G^{b]d}\gamma_d\epsilon_{1,i}+\frac{1}{8\alpha}\gamma_cG^{c[a}\gamma\cdot T_{ij}\gamma^{b]}\epsilon^{j}_1-\frac{i}{4\alpha}\gamma_cG^{c[a}\lambda_i\bar{\epsilon}_1\gamma^{b]}\lambda\nonumber\\
					&-\frac{1}{96}\gamma\cdot T_{ij}\gamma^{[a}G^{b]c}\gamma_c\epsilon^j_1+\frac{1}{8}\gamma\cdot T_{ij}\gamma^{[a}\gamma\cdot T^{jk}\gamma^{b]}\epsilon_{1,k}+\frac{i}{2}\gamma\cdot T_{ij}\gamma^{[a}\lambda^j\bar{\epsilon}_1\gamma^{b]}\lambda\nonumber\\
					&-\frac{i}{48}\lambda_i \bar{\lambda}\gamma^{[a}G^{b]c}\gamma_c\epsilon_1+\frac{i}{4}\lambda_i\bar{\lambda}^j\gamma^{[a}\gamma\cdot T_{jk}\gamma^{b]}\epsilon_1^k\bigg)+\frac{i}{24}\gamma^a\gamma_{ef}\bigg(\frac{i}{48}\gamma_c G^{ce}G^{fd}\gamma_d\epsilon_{1,i}\nonumber\\
					&+\frac{1}{8\alpha}\gamma_cG^{ce}\gamma\cdot T_{ij}\gamma^{f}\epsilon^j_{1}-\frac{i}{4\alpha}\gamma_cG^{ce}\lambda_i\bar{\epsilon}_1\gamma^{f}\lambda-\frac{1}{96}\gamma\cdot T_{ij}\gamma^{e}G^{fc}\gamma_c\epsilon^j_1\nonumber\\
					&+\frac{1}{8}\gamma\cdot T_{ij}\gamma^{e}\gamma\cdot T^{jk}\gamma^{f}\epsilon_{1,k}+\frac{i}{2}\gamma\cdot T_{ij}\gamma^{e}\lambda^j\bar{\epsilon}_1\gamma^{f}\lambda-\frac{i}{48}\lambda_i \bar{\lambda}\gamma^{e}G^{fc}\gamma_c\epsilon_1\nonumber\\
					&+ \frac{i}{4}\lambda_i\bar{\lambda}^j\gamma^{e}\gamma\cdot T_{jk}\gamma^{f}\epsilon_1^k\bigg)\bigg\}-(1\leftrightarrow 2)\bigg]\nonumber\\
					\Lambda_3^{(Q,Q)}{}^{ab} &= \frac{i}{4}\Bar{\epsilon}_2^i \gamma^{[a}\gamma_{cd}\gamma^{b]}\epsilon_1^jT^{cd}_{ij}-\frac{i}{4\alpha}\Bar{\epsilon}_2 \epsilon_1 G^{ab}-\frac{1}{8}(\Bar{\lambda}\gamma^{ab}\lambda\Bar{\epsilon}_2\epsilon_1+\Bar{\lambda}\gamma^{[a}\gamma_c\gamma^{b]} \lambda \Bar{\epsilon}_2\gamma^c\epsilon_1)\nonumber\\
					\Lambda^{(S,S)}_{a} &=\frac{1}{2}\Bar{\eta}_2 \gamma_a \eta_1\nonumber\\
					\eta^{(Q,S)}_3{}^i&=-\frac{10i}{3}\bar{\eta}^{(i}\epsilon^{j)}\lambda_{j}-\frac{i}{8}\bar{\epsilon}\gamma^{ab}\eta\gamma_{ab}\lambda^i-\frac{1}{16}\bar{\lambda}\gamma_{ab}\eta\gamma^{ab}\epsilon^i+\frac{i}{4}\bar{\eta}\gamma_a\lambda\gamma^a\epsilon^i+\frac{i}{4}\bar{\epsilon}\gamma^a\eta\gamma_a\lambda^i\nonumber\\
					&+\frac{1}{4}\bar{\epsilon}\gamma^a\lambda\gamma_a\eta^i\nonumber\\
					\Lambda_3^{(Q,S)}{}^{ij}&= -4i\Bar{\eta}^{(i} \epsilon^{j)}\nonumber\\
					\Lambda_3^{(Q,S)}{}^{ab} &=- \frac{i}{2}\Bar{\epsilon}\gamma^{ab}\eta\nonumber\\
					\Lambda^{(Q,S)}_{D3} &= -\frac{i}{2}\Bar{\epsilon}\eta\nonumber\\
					\Lambda_{3}^{(Q,S)}{}^a&=-\frac{i}{4}\bar{\eta}^i(\frac{i}{4\alpha}G^{ab}\gamma_b\epsilon_i+\frac{i}{4}T_{cd,ij}\gamma^{cd}\gamma^a\epsilon^j+\frac{1}{2}\bar{\epsilon}\gamma^a\lambda\lambda_i)
			\end{align}}
			
			\section{The $\mathcal{N}=2$ vector multiplet in five dimensions}
			
			In this section, we will obtain the vector multiplet for $\mathcal{N}=2$ conformal supergravity in five dimensions by dimensionally reducing the $(2,0)$ tensor multiplet in six dimensional conformal supergravity. As explained in section-\ref{6dtensor}, Table-\ref{table:2}, the $(2,0)$ tensor multiplet in six dimensions has the following fields: $B_{MN}, \psi^i, \phi^{ij}$.

			Below, we give the reduction ansatz of the fields of the six dimensional tensor multiplet to five dimensional fields. The fields appearing on the L.H.S are six dimensional fields whereas the fields appearing on the R.H.S are five dimensional fields and $\alpha$ is the scalar that appeared in the dimensional reduction of the six dimensional Vielbein.  
			{\allowdisplaybreaks
				\begin{align}\label{TensorVector}
					B_{\mu z}&=A_\mu\nonumber\\
					\psi^{i}&= \frac{\alpha}{4}\begin{pmatrix}
						0\\
						i\psi^i
					\end{pmatrix}\nonumber\\
					\phi^{ij}&= \alpha \sigma^{ij}\nonumber\\
			\end{align}}
			The anti-self dual part of the field strength of the six dimensional tensor gauge field $B_{MN}$ is constrained because of the equations of motion of the tensor multiplet in six dimensions as given in \ref{EOM}. This implies that $B_{\mu\nu}$ will not be an independent component and can be determined completely in terms of $B_{\mu z}$. Hence we do not include it in the reduction ansatz above. 
			
			Hence, the dimensional reduction of the six dimensional tensor multiplet on a circle gives rise to a vector multiplet in five dimensions whose details are given in table-\ref{table:4}.
			\begin{table}[h!]
				\centering
				\centering
				\begin{tabular}{ |p{1.5cm}|p{1.5cm}|p{7cm}|p{1cm}|p{1cm}|}
					\hline
					Field & Type& Properties & USp(4)&Weyl weight\\
					\hline
					$A_\mu$& Boson& real vector&\bf{1}&0\\
					$\psi^i$&Fermion&Symplectic Majorana spinor&\bf4&$\frac{3}{2}$\\
					$\sigma^{ij}$&Boson&$\sigma^{ij}=-\sigma^{ji}, \Omega_{ij}\sigma^{ij}=0$&\bf5&1\\
					\hline
				\end{tabular}
				\caption{The vector multiplet}
				\label{table:4}	
			\end{table}

			By using the transformation rules of the six dimensional tensor multiplet (eq-\ref{Tensor}), the dictionary (eq-\ref{TensorVector}) and the relation between the 6d and 5d supersymmetry transformations (\ref{6d-5d}), we obtain the following transformation rules for the five dimensional $\mathcal{N}=2$ vector multiplet coupled to conformal supergravity:  
			
			{\allowdisplaybreaks
				\begin{align}
					\delta A_\mu &=\frac{1}{4}\bar{\epsilon}\gamma_\mu\psi - \frac{1}{2}\bar{\epsilon^i}\gamma_\mu\lambda^j\sigma_{ij}+\frac{i}{2}\bar{\epsilon}^i\psi_\mu^j\sigma_{ij}\nonumber \\ 
					\delta\sigma^{ij}&=\Lambda_D \sigma^{ij}-\Lambda^{[i}{}_{k}\sigma^{j]k}-\frac{i}{2}\bar{\epsilon}\lambda\sigma^{ij}+i\bar{\epsilon}^{[i}\psi^{j]}-\Omega\text{trace}\nonumber\\
					\delta \psi^i &= \frac{3}{2}\Lambda_D \psi^{i}+\frac{1}{2}\Lambda^{i}{}_{j}\psi^{j}-\frac{1}{4}\Lambda^{ab}\gamma_{ab}\psi^{i}+\frac{1}{2} H. \gamma \epsilon^i-i\cancel{D}\sigma^{ij}\epsilon_j-\frac{i}{\alpha}\sigma^{ij}\cancel{D}\alpha\epsilon_j-\frac{i}{4}\bar{\epsilon}\gamma^{a}\lambda\gamma_{a}\psi^{i}\nonumber\\
					&-\frac{i}{8}\bar{\lambda}\gamma_{ab}\psi\gamma^{ab}\epsilon^i-\frac{i}{2}\bar{\epsilon}\lambda \psi^i+E^{[i}{}_k \sigma^{j]k}\epsilon_j-i\bigg(\bar{\lambda}^{[i}\psi^{j]}-\text{trace}\bigg)\epsilon_j-4\sigma^{ij}\eta_j\;,
				\end{align}
			}
			
			where $H_{\mu\nu}$ is the fully super-covariant field strength of the vector gauge field $A_{\mu}$ given as:
			\begin{align}
				H_{\mu \nu}&=2\partial_{[\mu}A_{\nu]}-\frac{1}{2}\bar{\psi}_{[\mu}\gamma_{\nu]}\psi+\bar{\psi}^i_{[\mu}\gamma_{\nu]}\lambda^j\sigma_{ij}-\frac{i}{2}\bar{\psi}^i_{\mu}\psi_\nu^j\sigma_{ij},
			\end{align}

			and 
			\begin{align}
				D_\mu\sigma^{ij}&=\partial_\mu\sigma^{ij}+\frac{i}{2}\bar{\psi}_\mu\lambda\sigma^{ij}-b_\mu\sigma^{ij}+V_\mu{}^{[i}{}_k\sigma^{j]k}-i\bigg(\bar{\psi}_\mu^{[i}\psi^{j]}-\Omega\text{trace}\bigg).\nonumber\\ 
			\end{align}

		We obtain the following field equations for the vector multiplet, by using the six dimensional field equations for the tensor multiplet (\ref{EOM}) and the reduction ansatz (eq- \ref{TensorVector}). The equations of motion also lead to the on-shell closure of the supersymmetry algebra on the vector multiplet.
		
			\begin{align}
				D_a \mathcal{H}_{ab}=&(^*\omega)_b\nonumber\\
				\cancel{{D}}\psi^i+\frac{1}{2\alpha}\cancel{{D}}\alpha \psi^i+\cancel{{D}}\sigma^{ij}\lambda_j+\frac{1}{5\alpha}\sigma^{ij}\cancel{{D}}\alpha \lambda_j+\frac{4}{5}\sigma^{ij}\cancel{{D}}\lambda_j+\frac{i}{8\alpha}G\cdot\gamma\psi^i&\nonumber\\
				+\frac{i}{5\alpha}\sigma^{ij}G\cdot\gamma\lambda_j+\frac{i}{2}E^i{}_j\psi^j+\frac{i}{2}H\cdot \gamma\lambda^i+\frac{i}{5}\sigma^{ij}T_{jk}\cdot\gamma\lambda^k+\frac{i}{2}\gamma \cdot T^{ij}\psi_j+iE^{[i}{}_k\sigma^{j]k}\lambda_j&\nonumber\\
				+\frac{i}{4}\sigma^{kl}\chi^i{}_{kl}+\bigg(\bar{\lambda}^{[i}\psi^{j]}-\text{trace}\bigg)\lambda_j+\frac{1}{8}\bar{\lambda}\gamma_{ab}\psi\gamma^{ab}\lambda^i=&0,\nonumber\\
				{D}^a{D}_a\sigma^{ij}+\frac{1}{\alpha}{D}^a\alpha{D}_a\sigma^{ij}-\frac{1}{5\alpha^2}\sigma^{ij} D^a\alpha D_a\alpha+\frac{3}{5\alpha}\sigma^{ij}{D}^a{D}_a\alpha&\nonumber\\
				-E^{k[i}\Omega^{j]l}\bigg\{E_{m[k}\sigma_{l]}{}^m-i\bigg(\bar{\lambda}_{[k}\psi_{l]}-\text{trace}\bigg)\bigg\}+\frac{i}{4}\bar{\lambda}_k\gamma\cdot T^{k[i}\psi^{j]}-\bigg(\bar{\lambda}_l\chi^{(i,k)l}\sigma^j{}_k-(i\leftrightarrow j)\bigg)&\nonumber\\
				-\bar{\lambda}^{[i}\bigg(-\frac{i}{8\alpha}{G}\cdot\gamma\psi^{j]}+\frac{1}{2\alpha}\cancel{{D}}\alpha\psi^{j]}-\frac{i}{2}E^{j]}{}_k\psi^k-\frac{i}{2}H\cdot\gamma\lambda^{j]}-\frac{1}{8}\bar{\lambda}\gamma_{ab}\psi\gamma^{ab}\lambda^{j]}-\cancel{{D}}\sigma^{j]k}\lambda_k&\nonumber\\
				-\frac{1}{5\alpha}\sigma^{j]k}\cancel{{D}}\alpha\lambda_k-\frac{4}{5}\sigma^{j]k}\cancel{{D}}\lambda_k+\frac{i}{5\alpha}\sigma^{j]k}G\cdot\gamma\lambda_k-\frac{i}{5}\sigma^{j]k}T_{kl}\cdot\gamma\lambda^l\bigg)+\bar{\lambda}^{[i}\Omega^{j]k}\bigg\{iE^m{}_{[k}\sigma_{l]m}\lambda^l&\nonumber\\
				-\bigg(\bar{\lambda}_{[k}\psi_{l]}-\text{trace}\bigg)\lambda^l\bigg\}+\frac{1}{5}\bar{\psi}^{[j}\bigg(\frac{1}{\alpha}\cancel{{D}}\alpha\lambda^{i]}-\cancel{{D}}\lambda^{i]}-\frac{i}{4\alpha}G \cdot\gamma\lambda^{i]}+\frac{i}{4}\gamma \cdot {T^{i]k}}\lambda_k                                                                                                                                                                                                                                                                                                                                                                \bigg)&\nonumber\\
				-\frac{1}{15}D^{kl,ij}\sigma_{kl}+6H\cdot T^{ij}-\frac{3i}{2}T^{ij}_{ab}\bar{\lambda}\gamma_{ab}\psi+\frac{i}{4}\chi^{k,ij}\psi_k-\frac{1}{10\alpha^2}\sigma^{ij}G_{ab}G^{ab}-\frac{1}{10}\sigma^{ij}T^{kl}_{ab}T_{kl}^{ab}=&0\nonumber\\
			\end{align}
			where 
			\begin{align}
				\mathcal{H}_{ab}&=\alpha H_{ab}-\frac{i\alpha}{4}\bar{\lambda}\gamma_{ab}\psi-\alpha\sigma_{ij}T^{ij}_{ab}\nonumber\\
				\omega_{abcd}&={-6G_{[ab}H_{cd]}-\frac{3i}{2}\alpha\bar{\psi}\gamma_{[ab}R(Q)_{cd]}}\nonumber\\
				(^*\omega)_a&=\frac{1}{4!}\epsilon_{abcde}\omega^{bcde}\nonumber\\
			\end{align}
			\section{Conclusion and Future Directions}
			In this paper, we have constructed five dimensional $\mathcal{N}=2$  conformal supergravity via dimensional reduction of six dimensional $(2,0)$ conformal supergravity on a circle. The dimensional reduction of $(2,0)$ Weyl multiplet in six dimensions gives us a dilaton weyl multiplet in five dimensions. We also derived the soft superconformal algebra that is realized on this multiplet. One can easily check that the rigid limit of this soft superconformal algebra is a super-Poincar{\'e} algebra and hence the existence of $\mathcal{N}=2$ conformal supergravity in five dimensions is not in disagreement with Nahm's classification which states that there cannot be a rigid $\mathcal{N}=2$ superconformal algebra in five dimensions. Typically, by construction, the soft superconformal algebra that is realized on a standard Weyl multiplet goes to a rigid superconformal algebra in the rigid limit. Nahm's classification regarding the non-existence of a rigid $\mathcal{N}=2$ superconformal algebra in five dimensions can only rule out the existence of a standard Weyl multiplet. But, as dilaton weyl multiplet has no direct connection with the rigid superconformal algebra, it cannot rule out the existence of a dilaton Weyl multiplet and hence conformal supergravity. Indeed via dimensional reduction of $(2,0)$ conformal supergravity in six dimensions we have a $\mathcal{N}=2$ conformal supergravity in five dimensions with the dilaton Weyl multiplet instead of a standard Weyl multiplet.

			Matter multiplets play a crucial role in the construction of Poincar{\'e} supergravity from conformal supergravity using the methods of supermultiplet calculus. Hence, we also constructed the five dimensional $\mathcal{N}=2$ on-shell vector multiplet by dimensionally reducing the six dimensional  $(2,0)$ on-shell tensor multiplet.
			
			In future, we would like to develop an action density formula for $\mathcal{N}=2$ conformal supergravity in five dimensions following the general prescription discussed in \cite{Butter:2019edc,Hegde:2021rte,Hegde:2019ioy}. Further, this density formula can be used in constructing the most general action for the Weyl multiplet as well as the vector multiplets. This can then be used for constructing the most general matter coupled $\mathcal{N}=2$ Poincar{\'e} supergravity in five dimensions with higher derivative corrections. 
			
			We would also like to generalize the vector multiplet constructed in this paper to a Yang-Mills multiplet. It has been seen in the case of six dimensional $(1,0)$ \cite{Bergshoeff:1986vy}, five dimensional $\mathcal{N}=1$ \cite{Ozkan:2013nwa,Ozkan:2013uk} and four dimensional $\mathcal{N}=2$ \cite{Mishra:2020jlc} supergravity that upon Poincar{\'e} gauge fixing there is a mapping between the dilaton Weyl multiplet and the Yang-Mills multiplet which leads to the supersymmetrization of the Riemann squared term in the respective Poincar{\'e} supergravity. We would also like to attempt such a construction for five dimensional $\mathcal{N}=2$ Poincar{\'e} supergravity.

			In the literature, conformal supergravity has been constructed primarily on those cases where the background flat rigid superconformal algebra is known and given in Nahm's classification. But, similar to our scenario, where no known algebra exists, conformal supergravity had been constructed earlier, for example $\mathcal{N}=1$ conformal supergravity in ten dimensions \cite{Bergshoeff:1982az}. We would like to explore this not-so-well-known aspect of conformal supergravity.
			
			\acknowledgments
			We would like to thank Subrabalan Murugesan for his involvement in the initial stage of the project. We would like to thank Daniel Butter for useful discussions. We would like to thank Subramanya Hegde for his help in checking the Gamma Matrix identities using Mathematica and his comments on the draft. We would also like to thank Madhu Mishra for useful discussions and her comments on the draft. SA thanks IMSc for hospitality during the course of this work. This work has been partially supported by SERB core research grant CRG/2018/002373, Government of India.
			
			\appendix
			\section{Conventions and notations}\label{conv}
				The convention for metric that we follow is mostly positive. The bar on any spinor is always a Majorana conjugate defined as: 
			\begin{align}\label{Majconj}
				\bar{\lambda}^i=(\lambda^i)^T C\;,
			\end{align}
			where $C$ is the charge conjugation matrix. In five dimension, an irreducible spinor is symplectic Majorana which has 8 degrees of freedom. An $USp(4)$ symplectic Majorana spinor which is relevant for the paper has 16 components encoded in a quartet $\chi^{i}$, where $i=1,2,3,4$.  The symplectic Majorana condition on $\chi^{i}$ is given as follows: 
			\begin{equation}\label{sympMaj}
				\bar{\chi}^i=-i\Omega^{ij}(\chi^j)^{\dagger}\gamma^0
			\end{equation}
			where $\Omega^{ij}$ is a real anti-symmetric tensor which is an $USp(4)$ invariant\footnote{If the spinor is an $SU(2)$ symplectic Majoarana spinor, as is the case for $\mathcal{N}=1$ conformal supergravity, then $\Omega^{ij}$ would be replaced by $\varepsilon^{ij}$ where $\varepsilon^{ij}$ is the Levi-Civita tensor and is an $SU(2)$ invariant tensor. However, in this paper we will restrict ourselves to $USp(4)$ symplectic Majorana spinors and hence we will use $\Omega^{ij}$ which is an $USp(4)$ invariant.}. Its inverse $\Omega_{ij}$ is defined via the condition $\Omega_{ij}\Omega^{ik}=\delta_j^{~k}$.
			
			An irreducible six dimensional spinor is also a symplectic Majorana spinor satisfying the relation
			\begin{equation}\label{sympMaj6d}
				\bar{\chi}^i=i\Omega^{ij}(\chi^j)^{\dagger}\Gamma^0
			\end{equation}
			Additionally, a chirality constraint can be imposed on the spinor which is either: $\Gamma_* \chi^i=\chi^i$ or  $\Gamma_* \chi^i=-\chi^i$, where $\Gamma_*=\Gamma^{0}\Gamma^{1}\Gamma^{2}\Gamma^{3}\Gamma^{4}\Gamma^{5}$ is the chirality matrix in six dimensions. When the first condition is satisfied, the spinor is called a left chiral symplectic Majorana spinor and when the second condition is satisfied, the spinor is called a right chiral symplectic Majorana spinor.

			We can raise and lower the $USp(4)$ index of a spinor by using the $USp(4)$ invariant $\Omega_{ij}$: $\chi_i=\chi^j\Omega_{ji}$ and similarly $\chi^i=\Omega^{ij}\chi_j$.
			
			When the $USp(4)$ indices are omitted in the bilinears, northwest-southeast contraction is being implied, for e.g. $\bar{\chi}\gamma^{(r)}\lambda=\bar{\chi}^i\gamma^{(r)}\lambda_i$, where the $\gamma^{(r)}$ is a rank-$r$ element of the Clifford algebra defined as follows: $\gamma^{(r)}=\gamma^{a_1 a_2...a_r}=\gamma^{[a_1}\gamma^{a_2}...\gamma^{a_r]}$.

			The order of the spinors in a bilinear is changed in the following way:
			
			\begin{equation}\label{orderchange}
				\bar{\chi}\gamma^{(r)} \lambda=t_r \bar{\lambda}\gamma^{(r)} \chi\;,
			\end{equation}
			where $t_r=\pm1$ is a property of the rank-$r$ and the space-time dimension and it satisfies $t_{r+4}=t_r$. For five dimensions, $t_0=t_1=1$ and $t_2=t_3=-1$. For six dimensions, $t_1=t_2=1$ and $t_0=t_3=-1$.
			
			The Fierz re-arrangement identity in five dimensions, which we often use, is given as:
			$\psi_j\bar{\lambda}^i=-\frac{1}{4}\bar{\lambda}^i\psi_j-\frac{1}{4}\bar{\lambda}^i\gamma^a\psi_j\gamma_a+\frac{1}{8}\bar{\lambda}^i\gamma^{ab}\psi_j\gamma_{ab}.$
			
			\section{Dimensional Reduction of spinors and gamma matrices}\label{spin_Gamma_red}
				Here, we give the necessary relations that are needed in dimensional reduction scheme for spinors.
			
			Let us consider a left chiral symplectic Majorana spinor $\lambda_1$ and a right chiral symplectic Majorana spinor $\lambda_2$ in six dimensions. In our convention of Clifford algebra given in (\ref{Gamma6d}, \ref{Gamma7}, \ref{C6d}), they reduce to five dimensional symplectic Majorana spinors $\psi$ and $\sigma$ as follows:
			\begin{align}
				\lambda_1&=\begin{pmatrix}
					\psi \\
					0
				\end{pmatrix},\nonumber\\
				\lambda_2 &=\begin{pmatrix}
					0 \\
					i\sigma
				\end{pmatrix}  
			\end{align}
			
			The Majorana conjugate of the above six dimensional spinors reduces to the five dimensional ones in the following way: 
			\begin{align}
				\bar{\lambda}_1&=\begin{pmatrix}
					0 & -\bar{\psi}
				\end{pmatrix},\nonumber\\
				\bar{\lambda}_2&=\begin{pmatrix}
					i\bar{\sigma} & 0
				\end{pmatrix}.
			\end{align}

			We also give the decomposition of the elements of Clifford algebra in six dimensions into the Clifford algebra elements of five dimensions up to rank three.
			\begin{align}\label{Gamma6d}
				\Gamma_A&=\begin{cases}
					\begin{pmatrix}
						0 &\gamma_a \\
						\gamma_a & 0
					\end{pmatrix}
					& \text{for}\ A=a\leq4 \\
					\begin{pmatrix}
						0 &-i\boldsymbol{1}_4 \\
						i\boldsymbol{1}_4  & 0
					\end{pmatrix}               & A=5
				\end{cases}
				,\nonumber\\
				\Gamma_{AB}&=\begin{cases}
					\begin{pmatrix}
						\gamma_{ab} &0 \\
						0 & \gamma_{ab}
					\end{pmatrix}
					& \text{for}\ A,B=a,b\leq4 \\
					\begin{pmatrix}
						i\gamma_a & 0 \\
						0  & -i\gamma_a
					\end{pmatrix}               & A=a \leq4 ~\&~ B=5
				\end{cases} 
				,\nonumber\\
				\Gamma_{ABC}&=\begin{cases}
					\begin{pmatrix}
						0 & \gamma_{abc}\\
						\gamma_{abc} & 0
					\end{pmatrix}
					& \text{for}\ A,B,C=a,b,c\leq4 \\
					\begin{pmatrix}
						0 & -i\gamma_{ab} \\
						i\gamma_{ab} & 0
					\end{pmatrix}               & A,B=a,b \leq4 ~\&~ C=5
				\end{cases} 
			\end{align}
			
			The six dimensional chirality matrix reduces as follows: 
			\begin{align}\label{Gamma7}
				\Gamma_*=\begin{pmatrix}
					\boldsymbol{1}_4 & 0 \\
					0 & -\boldsymbol{1}_4
				\end{pmatrix}
			\end{align}

			Six dimensional charge conjugation matrix is decomposed into the five dimenisional ocnjugation matrix by the following relation: 
			\begin{align}\label{C6d}
				C_{6D}=\begin{pmatrix}
					0 & -C_{5D}\\
					C_{5D} & 0
				\end{pmatrix}
			\end{align}
			\section{Gamma matrices properties}
			\begin{align}\label{cliffalg}
				[\gamma_a , \gamma_{bc}] &=4\eta_{a[b} \gamma_{c]}, 	\{ \gamma_{ab} , \gamma_c \} = 2\gamma_{abc};\nonumber\\
				\gamma^b \gamma_a \gamma_b &= -3\gamma_a , \gamma^{ab}\gamma_{ab} = 20;\nonumber\\
				\gamma^c \gamma_{ab} \gamma_c &= \gamma_{ab},
				\gamma_{cd} \gamma_a \gamma^{cd} = -4\gamma_a, \gamma^{cd}\gamma_{ab}\gamma_{cd}=4\gamma_{ab};\nonumber\\
				[\gamma_{ab}, \gamma^{cd}] &=-8\delta_{[a}^{~[c} \gamma_{b]}^{~~d]},\{\gamma^{ab},\gamma^{cd}\}=-4\eta^{a[c}\eta^{d]b}+2\gamma^{abcd}.\nonumber\\
				\gamma^{ab}&=\frac{-i}{3!}\epsilon^{abcde}\gamma_{cde}\nonumber\\
				\gamma^{abc}&=-\frac{i}{2!}\epsilon^{abcde}\gamma_{de}\nonumber\\
				\gamma^{abcd} &=i\epsilon^{abcde}\gamma_e\nonumber\\
				\end{align} 
			
			\section{Dependent gauge fields transformations and the corresponding curvatures}
				For the sake of completeness, we give the gauge transformations of the dependent gauge fields both in six as well as five dimensions. 
			\subsection{Six-dimensional dependent gauge fields}
				The gauge transformations of the six dimensional dependent gauge fields are as follows
			{\allowdisplaybreaks
				\begin{align}\label{6ddep}
					\delta\omega_{M}{}^{AB}&=\partial_{M}\Lambda^{AB}-4e_{M}{}^{[A}\Lambda^{B]}-2\Lambda^{[A}{}_C \omega_M{}^{B]C}-\frac{1}{2}\bar{\epsilon}\Gamma^{AB}\phi_{M}-\frac{1}{2}\bar{\epsilon}\Gamma^{[A}R(Q)_M^{~~B]}\nonumber\\
					&-\frac{1}{4}\bar{\epsilon}\Gamma_M R(Q)^{AB}-\frac{1}{2}\bar{\epsilon}_i\Gamma_C\psi_{M,j}T^{ABC,ij}\nonumber\\
					\delta \phi_{M}^{i}&=D_{M}\eta^{i}+\frac{1}{2}\Lambda \phi_{M}^{i}+\frac{1}{2}\Lambda^{i}{}_{j}\phi_{M}^{j}-\frac{1}{4}\Lambda^{AB}\Gamma_{AB}\phi^i_M+\frac{1}{6}\bigg\{-\frac{1}{8}\bar{\psi}_M\Gamma^{[A}R(Q)_C^{~~B]}\Gamma^C \Gamma_{AB}\epsilon^i\nonumber\\
					&-\frac{1}{16}\bar{\psi}_M\Gamma_C R(Q)^{AB}\Gamma^C \Gamma_{AB}\epsilon^i-\frac{2}{15}\bar{\psi}_{M,k}\Gamma_C\chi^{(i,j)k}\Gamma^C\epsilon_j+\frac{1}{12}\bigg(-\bar{\epsilon}{}^{[i}R(Q)_{AB}{}^{j]} \nonumber\\
					&-\frac{1}{15}\bar{\epsilon}{}^k\Gamma_{AB}\chi_k{}^{ij}\bigg)\Gamma^{AB}\psi_{M,j}-\Omega\text{trace}\bigg\}-\frac{1}{6}\bigg\{-\frac{1}{8}\bar{\epsilon}\Gamma^{[A}R(Q)_C{}^{B]}\Gamma^C \Gamma_{AB}\psi_M^i\nonumber\\
					&-\frac{1}{16}\bar{\epsilon}\Gamma_C R(Q)^{AB}\Gamma^C \Gamma_{AB}\psi_M^i-\frac{2}{15}\bar{\epsilon}_k\Gamma_C\chi^{(i,j)k}\Gamma^C\psi_{M,j}+\frac{1}{12 }\bigg(-\bar{\psi}_{M}{}^{[i}R(Q)_{AB}{}^{j]}\nonumber\\
					&-\frac{1}{15}\bar{\psi}_M{}^k\Gamma_{AB}\chi_k{}^{ij}\bigg)\Gamma^{AB}\epsilon_{j}-\Omega\text{trace}\bigg\}-\frac{1}{4}\bigg\{\Gamma^C\bigg(-\frac{1}{4}R(M)_{MC}{}^{AB}\Gamma_{AB}\epsilon^i\nonumber\\
					&+\frac{1}{2}R(V)_{MC}^{}{i}{}_{j}\epsilon^j-\frac{1}{168}e_M{}^D\Gamma\cdot T^{ij}\Gamma_{[C}\Gamma \cdot T_{jk}\Gamma_{D]}\epsilon^k\bigg)+\frac{\Gamma_M}{10}\Gamma^{CD}\bigg(\frac{1}{4}R(M)_{CD}{}^{AB}\Gamma_{AB}\epsilon^i\nonumber\\
					&-\frac{1}{2}R(V)_{CD}{}^{i}{}_{j}\epsilon^j-\frac{1}{168}\Gamma \cdot T^{ij}\Gamma_C\Gamma\cdot T_{jk}\Gamma_D\epsilon^k\bigg)\bigg\}+f_M^A\Gamma_A\epsilon^i -\Gamma_A \Lambda^A\psi_\mu^i\nonumber\\
					\delta f_{M}{}^{A}&=\partial_{M}\Lambda^{A}+\Lambda f_{M}{}^{A}-\Lambda^{AB}f_{M B}-\frac{1}{16}\bar{\epsilon}^i\bigg[-\frac{1}{2}R(M)^{ABCD}\Gamma_{BCD}\psi_{M,i}+\Gamma_BR(V)^{AB}{}_{ij}\psi_M^j\nonumber\\
					&-\frac{1}{10}\bigg\{\frac{1}{4}\bigg(\Gamma^A\Gamma^{BC}\Gamma^{DE}+\Gamma^{DE}\Gamma^{BC}\Gamma^A\bigg)R(M)_{BC,DE}\psi_{M,i}-R(V)^{BC}{}_{ij}\Gamma^{ABC}\psi^j_M\bigg\}\bigg]\nonumber\\
					&+\frac{1}{4}\bar{\phi}^i_M T_{ij}\cdot 
					\Gamma\Gamma^A \epsilon^j-\frac{1}{16}\bar{\epsilon}^i\Gamma_B R(Q)_{CM}^jT^{ABC}_{ij}-\frac{1}{16}\bar{\epsilon}\Gamma^{AB}R(S)_{MB}\nonumber\\
					&+\frac{1}{160}e_M{}^A\bigg(\bar{\epsilon}^i\Gamma_FR(Q)_{DE}^jT^{DEF}_{ij}+\bar{\epsilon}\Gamma^{CD}R(S)_{CD}\bigg)\nonumber\\
					&+\bigg(\frac{1}{384}\bar{R}(Q)^i_{M}{}^{[A}\Gamma^{B]}\Gamma\cdot T_{ij}\Gamma_{B}\epsilon^j+\frac{1}{32}\bar{R}(Q)^{AB,i}\Gamma_{M}\Gamma\cdot T_{ij}\Gamma_{B}\epsilon^j\bigg)\nonumber\\
					&+\frac{1}{32}\bigg(\frac{1}{8}\bar{\epsilon}^{[i}\Gamma^{DE}\Gamma_{MBC}R_{DE}^{j]}(Q)-\frac{1}{15}\bar{\epsilon}^k\Gamma_{MBC}\chi_k{}^{ij}-\Omega\text{trace}\bigg)T_{ij}^{ABC}\nonumber\\
					&+\frac{1}{32}T_{MBC,ij}\bigg(\frac{1}{8}\bar{\epsilon}^{[i}\Gamma^{DE}\Gamma^{ABC}R_{DE}^{j]}(Q)-\frac{1}{15}\bar{\epsilon}^k\Gamma^{ABC}\chi_k{}^{ij}-\Omega\text{trace}\bigg)\nonumber\\
					&-\frac{1}{160}e_M^AT^{BCF}_{ij}\bigg(\frac{1}{8}\bar{\epsilon}^{[i}\Gamma^{DE}\Gamma_{BCF}R_{DE}^{j]}(Q)-\frac{1}{15}\bar{\epsilon}^k\Gamma_{BCF}\chi_k{}^{ij}-\Omega\text{trace}\bigg)\nonumber\\
					&-\frac{1}{4}\bar{\eta}^i T_{ij}\cdot\Gamma \Gamma^A\psi_M^j + \frac{1}{2}\bar{\phi}_M \gamma^A\eta\;,\nonumber\\
						\end{align}}

			where 
			\begin{align}
				\mathcal{D}_{M} \eta^i =\partial_{M} \eta^i-\frac{1}{2}b_M \eta^i-\frac{1}{2}V_M^{i}{}_{j}\eta^{j}+\frac{1}{4}\omega_M^{AB}\Gamma_{AB}\eta^i
			\end{align}
		
		The corresponding curvatures are given below:
		\begin{align}
			R(S)_{MN}^i&=2\partial_{[M}\phi_{N]}{}^i-b_{[M} \phi_{N]}^{i}+V_{[M}^{ij}\phi_{N],j}+\frac{1}{2}\omega_{[M}^{ab}\Gamma_{ab}\phi^i_{N]}+2f_{[M}^A\Gamma_A\psi_{N]}^i\nonumber\\
			&+\frac{1}{3}\bigg\{-\frac{1}{8}\bar{\psi}_{[M}\Gamma^{[A}R(Q)_C^{~~B]}\Gamma^C \Gamma_{AB}\psi_{N]}^i-\frac{1}{16}\bar{\psi}_{[M}\Gamma_C R(Q)^{AB}\Gamma^C \Gamma_{AB}\psi_{N]}^i\nonumber\\
			&+\frac{2}{15}\bar{\psi}_{[M}^k\Gamma_C\chi^{(i,j)}{}_k\Gamma^C\psi_{N],j}+\frac{1}{12}\bigg(-\bar{\psi}_{[N}^{[i}R(Q)_{AB}{}^{j]}-\frac{1}{15}\bar{\psi}_{[N}^k\Gamma_{AB}\chi_k{}^{ij}\bigg)\Gamma^{AB}\psi_{M],j} \nonumber\\
			&-\Omega\text{trace}\bigg\}-\frac{1}{2}\bigg\{\Gamma^C\bigg(\frac{1}{4}R(M)_{C[M}{}^{AB}\Gamma_{AB}\psi_{N]}^i+\frac{1}{2}R(V)_{C[M}{}^{i}{}_{j}\psi_{N]}^j\nonumber\\
			&-\frac{1}{168}\Gamma\cdot T^{ij}\Gamma_{[C}\Gamma \cdot T_{jk}\Gamma_{D]}e_{[M}^D\psi_{N]}^k\bigg)+\frac{1}{10}\Gamma_{[M}\Gamma^{CD}\bigg(\frac{1}{4}R(M)_{CD}{}^{AB}\Gamma_{AB}\psi_{N]}^i\nonumber\\
			&-\frac{1}{2}R(V)_{CD}{}^{i}{}_{j}\psi_{N]}^j-\frac{1}{168}\Gamma \cdot T^{ij}\Gamma_C\Gamma\cdot T_{jk}\Gamma_D\psi_{N]}^k\bigg)\bigg\} \nonumber\\
			R(K)_{MN}{}^A&=2\partial_{[M} f_{N]}{}^A-2b_{[M}f_{N]}{}^A-2f_{B[N }\omega_{M]}{}^{AB}-\frac{1}{16}\bar{\psi}_{[N}^i\bigg[-\frac{1}{2}R(M)^{ABCD}\Gamma_{BCD}\psi_{M],i}\nonumber\\
			&+\Gamma_BR(V)^{AB}{}_{ij}\psi_{M]}^j-\frac{1}{10}\bigg\{\frac{1}{4}\bigg(\Gamma^A\Gamma^{BC}\Gamma^{DE}+\Gamma^{DE}\Gamma^{BC}\Gamma^A\bigg)R(M)_{BC,DE}\psi_{M],i}\nonumber\\
			&-R(V)^{BC}{}_{ij}\Gamma^{ABC}\psi^j_{M]}\bigg\}\bigg]+\frac{1}{2}\bar{\phi}^i_{[M} T_{ij}\cdot 
			\Gamma\Gamma^A \psi_{N]}^j+\frac{1}{8}\bar{\psi}_{[N}^i\Gamma_B R(Q)_{M]C}^jT^{ABC}_{ij}\nonumber\\
			&-\frac{1}{8}\bar{\psi}_{[N}\Gamma^{AB}R(S)_{M]B}+\frac{1}{160}e_{[M}^A\bigg(\bar{\psi}_{N]}^i\Gamma_FR(Q)_{DE}^jT^{DEF}_{ij}+\bar{\psi}_{N]}\Gamma^{CD}R(S)_{CD}\bigg)\nonumber\\
			&+\bigg(\frac{1}{384}\bar{R}(Q)^i_{[M}{}^{[A}\Gamma^{B]}\Gamma\cdot T_{ij}\Gamma_{B}\psi_{N]}^j+\frac{1}{32}\bar{R}(Q)^{AB,i}\Gamma_{[M}\Gamma\cdot T_{ij}\Gamma_{B}\psi_{N]}^j\bigg)\nonumber\\
			& + \frac{1}{2}\bar{\phi}_{[M} \gamma^A\phi_{N]}\;,\nonumber\\
		\end{align} 
			\subsection{Five-dimensional dependent gauge fields}
			The gauge transformations of the five dimensional dependent gauge fields are as follows
			{\allowdisplaybreaks
				\begin{align}
					\delta \omega_\mu ^{ab}  &=\partial_{\mu}\Lambda^{ab}-2\Lambda^{[a}{}_c\omega_\mu{}^{b]c}-4e_{\mu}^{[a}\Lambda^{b]} +\frac{i}{4}\Bar{\epsilon}^i \gamma^{[a}\gamma_{cd}\gamma^{b]}\psi_\mu^jT^{cd}_{ij}-\frac{i}{4\alpha}\Bar{\epsilon} \psi_\mu G^{ab}-\frac{1}{8}[\Bar{\lambda}\gamma^{ab}\lambda\Bar{\epsilon}\psi_\mu\nonumber\\
					&+\Bar{\lambda}\gamma^{[a}\gamma_c\gamma^{b]} \lambda \Bar{\epsilon}\gamma^c\psi_\mu]+\frac{i}{2}\Bar{\epsilon}\gamma^{ab}\phi_\mu +\frac{1}{4}\bar{\epsilon}\gamma_\mu R(Q)^{ab}+\frac{1}{2}\bar{\epsilon}\gamma^{[a}R(Q)_\mu{}^{b]}-\frac{i}{2}\bar{\eta}\gamma^{ab}\psi_\mu \nonumber\\
					\delta \phi_\mu^i&=D_{\mu}\eta^{i}+\frac{1}{2}\Lambda \phi_{\mu}^{i}+\frac{1}{2}\Lambda^{i}{}_{j}\phi_{\mu}^{j}-\frac{1}{4}\Lambda^{ab}\gamma_{ab}\phi^i_\mu -\biggl[\gamma^{ab}\epsilon^i\bigg(-\frac{i}{16}\bar{\psi}_\mu R(Q)_{ab}+\frac{1}{8}\bar{\psi}_\mu\gamma_a D_b\lambda\nonumber \\
					&-\frac{1}{8\alpha}D_a\alpha \bar{\psi}_\mu\gamma_b\lambda+\frac{i}{32\alpha}\bar{\lambda}\gamma_a\gamma_c G_{bc}\psi_\mu+\frac{i}{32}\bar{\lambda}^j\gamma_a\gamma\cdot T_{jk}\gamma_b\psi_\mu^k\bigg)+\gamma^a\epsilon^i\bigg(\frac{1}{4}\bar{\psi}_\mu D_a\lambda\nonumber\\
					&+\frac{i}{16\alpha}\bar{\lambda}G_{ac}\gamma^c\psi_{\mu}+\frac{i}{16}\bar{\lambda}^j\gamma\cdot T_{jk}\gamma_a\psi_{\mu}^k\bigg)+\frac{1}{4}\gamma^{ab}\epsilon_{j}\biggl\{\frac{i}{8\alpha}\bar{\psi}_\mu^{[i}\gamma \cdot G\gamma_{ab}\lambda^{j]}+\frac{1}{4}\bar{\psi}_\mu^{[i}\gamma^c\gamma_{ab}\nonumber\\
					&\bigg(-\frac{1}{\alpha}D_c\alpha\lambda^{j]}+D_c\lambda^{j]}+\frac{i}{4\alpha}G_{cd}\gamma^d\lambda^{j]}-\frac{i}{4}T^{j]k}\cdot\gamma \gamma_c\lambda_k\bigg)+\frac{1}{20}\bar{\psi}_\mu^{[i}\gamma_{ab}\bigg(\frac{1}{\alpha}\cancel{D}\alpha\lambda^{j]}-\cancel{D}\lambda^{j]}\nonumber\\
					&-\frac{i}{4\alpha}G\cdot\gamma \lambda^{j]}+\frac{i}{4}T^{j]k}\cdot\gamma\lambda_k\bigg)-\frac{i}{16}\bar{\psi}_\mu^k\gamma_{ab}\chi_k^{ij}-\frac{i}{8}\bar{\psi}_\mu^{[i}\gamma^{cd}\gamma_{ab}R(Q)_{cd}^{j]}\nonumber\\
					&+\frac{i}{4}\bar{\psi}_\mu\gamma_{abcd}\lambda T^{cd,ij}-\Omega\text{trace}\biggr\}-\frac{i}{4}\bar{\epsilon}\gamma^a\lambda\gamma_a\bigg(-\frac{1}{8\alpha} G \cdot \gamma \psi_\mu^i 
					+ \frac{i}{2\alpha} \cancel{D}\alpha\psi_\mu^i-\frac{1}{2}E^i_{~j}\psi_\mu^j 
					 \nonumber\\
					&+\frac{1}{4}\gamma \cdot T^{ij}\psi_{\mu,j} 
					+\frac{i}{4}\bar{\psi}_\mu\gamma ^a \lambda \gamma _a \lambda^i+\frac{i}{2}\bar{\psi}_\mu\lambda\lambda^i\bigg)+\frac{1}{2}\epsilon_{j}\bigg(-\frac{4}{5\alpha}\bar{\psi}_\mu^{(i} \lambda^{j)} \cancel{D}\alpha 
				\nonumber\\
					&	+\frac{4}{5}\bar{\psi}_\mu^{(i}\cancel{D}\lambda^{j)}-i\bar{\psi}_\mu^{(i}T^{j)k} \cdot \gamma \lambda_k
					-\frac{i}{4}\bar{\psi}_{\mu,k}\chi^{(i,j)k}
					-\frac{i}{2}E^{ij}\bar{\psi}_\mu\lambda\bigg)-\frac{i}{2}\bar{\epsilon}\lambda\bigg(-\frac{1}{8\alpha}\gamma \cdot G \psi_\mu^i 
				\nonumber\\
					&	+ \frac{i}{2\alpha} \cancel{D} \alpha \psi_\mu^i 
					-\frac{1}{2}E^i_{~j}\psi_\mu^j+\frac{1}{4}\gamma\cdot T^{ij}\psi_{\mu,j}+\frac{i}{2}\bar{\psi}_\mu\lambda\lambda^i 
					+\frac{i}{2}\bar{\psi}_\mu\gamma ^a \lambda \gamma _a \lambda^i\bigg)\nonumber\\
					&-\gamma_a\lambda^i\bigg(-\frac{1}{16}\bar{\psi}_\mu\gamma_b\lambda\bar{\epsilon}\gamma^a\gamma^b\lambda-\frac{1}{8}\bar{\psi}_\mu\lambda\bar{\epsilon}\gamma^a\lambda\bigg)\biggr]-\gamma_af^a_\mu\epsilon^i\nonumber\\
					&+ \biggl[\gamma^{ab}\psi_\mu^i\bigg(-\frac{i}{16}\bar{\epsilon}R(Q)_{ab}+\frac{1}{8}\bar{\epsilon}\gamma_a D_b\lambda-\frac{1}{8\alpha}D_a\alpha \bar{\epsilon}\gamma_b\lambda+\frac{i}{32\alpha}\bar{\lambda}\gamma_a\gamma_c G_{bc}\epsilon\nonumber\\
					&+\frac{i}{32}\bar{\lambda}^j\gamma_a\gamma\cdot T_{jk}\gamma_b\epsilon^k\bigg)+\gamma^a\psi_\mu^i\bigg(\frac{1}{4}\bar{\epsilon}D_a\lambda+\frac{i}{16\alpha}\bar{\lambda}G_{ac}\gamma^c\epsilon+\frac{i}{16}\bar{\lambda}^j\gamma\cdot T_{jk}\gamma_a\epsilon^k\bigg)\nonumber\\
					&+\frac{1}{4}\gamma^{ab}\psi_{\mu,j}\biggl\{\frac{i}{8\alpha}\bar{\epsilon}^{[i}\gamma \cdot G\gamma_{ab}\lambda^{j]}+\frac{1}{4}\bar{\epsilon}^{[i}\gamma^c\gamma_{ab}\bigg(-\frac{1}{\alpha}D_c\alpha\lambda^{j]}+D_c\lambda^{j]}+\frac{i}{4\alpha}G_{cd}\gamma^d\lambda^{j]}\nonumber\\
					&-\frac{i}{4}T^{j]k}\cdot\gamma \gamma_c\lambda_k\bigg)+\frac{1}{20}\bar{\epsilon}^{[i}\gamma_{ab}\bigg(\frac{1}{\alpha}\cancel{D}\alpha\lambda^{j]}-\cancel{D}\lambda^{j]}-\frac{i}{4\alpha}G\cdot\gamma \lambda^{j]}+\frac{i}{4}T^{j]k}\cdot\gamma\lambda_k\bigg)\nonumber\\
					&-\frac{i}{16}\bar{\epsilon}^k\gamma_{ab}\chi_k^{ij}+\frac{i}{4}\bar{\epsilon}\gamma_{abcd}\lambda T^{cd,ij}-\frac{i}{8}\bar{\epsilon}^{[i}\gamma^{cd}\gamma_{ab}R(Q)_{cd}^{j]}-\Omega\text{trace}\biggr\}\nonumber\\
					&-\frac{i}{4}\bar{\psi}_\mu\gamma^a\lambda\gamma_a\bigg(-\frac{1}{8\alpha} G \cdot \gamma \epsilon^i -\frac{1}{2}E^i_{~j}\epsilon^j 
					+\frac{1}{4}\gamma\cdot T^{ij} \epsilon_{j} 
					+\frac{i}{4}\bar{\epsilon}\gamma ^a \lambda \gamma _a \lambda^i
					+ \frac{i}{2\alpha} D^a \alpha \cdot \gamma_a \epsilon^i \nonumber\\
					&+\frac{i}{2}\bar{\epsilon}\lambda\lambda^i\bigg)+\frac{1}{2}\psi_{\mu,j}\bigg(-\frac{4}{5\alpha}\bar{\epsilon}^{(i} \lambda^{j)} \cancel{D}\alpha 
					+\frac{4}{5}\bar{\epsilon}^{(i}\cancel{D}\lambda^{j)}-i\bar{\epsilon}^{(i}T^{j)k} \cdot \gamma \lambda_k
					-\frac{i}{4}\bar{\epsilon}_{k}\chi^{(i,j)k}
					-\frac{i}{2}E^{ij}\bar{\epsilon}\lambda\bigg)\nonumber\\
					&-\frac{i}{2}\bar{\psi}_\mu\lambda\bigg(-\frac{1}{8\alpha}\gamma \cdot G \epsilon^i 
					+ \frac{i}{2\alpha} \cancel{D} \alpha \epsilon^i 
					-\frac{1}{2}E^i_{~j}\epsilon^j+\frac{1}{4}\gamma\cdot T^{ij}\epsilon_{j}\nonumber\\
					&+\frac{i}{2}\bar{\epsilon}\lambda\lambda^i 
					+\frac{i}{2}\bar{\epsilon}\gamma ^a \lambda \gamma _a \lambda^i\bigg)-\gamma_a\lambda^i\bigg(-\frac{1}{16}\bar{\epsilon}\gamma_b\lambda\bar{\psi}_\mu\gamma^a\gamma^b\lambda-\frac{1}{8}\bar{\epsilon}\lambda\bar{\psi}_\mu\gamma^a\lambda\bigg)\biggr]\nonumber\\
					&+i\lambda\bigg(\frac{i}{2\alpha}\bar{\epsilon}\cancel{D}\alpha\psi_\mu-\frac{1}{4}\bar{\epsilon}^j\gamma\cdot T_{jk}\psi_\mu^k\bigg)-\gamma_b\lambda^i\bigg(\frac{i}{8\alpha}\bar{\epsilon}\gamma_a\psi_{\mu}G^{ab}+\frac{1}{4\alpha}\bar{\epsilon}D^b\alpha\psi_{\mu}\nonumber\\
					&+\frac{i}{4}\bar{\epsilon}^j\gamma_a\psi_\mu^k T^{ab}_{jk}\bigg)-\frac{i}{32\alpha}(G\cdot\gamma-4i\cancel{D}\alpha)(2\Bar{\psi}_\mu \epsilon \lambda^i+\Bar{\psi}_\mu\gamma_a \lambda \gamma^a \epsilon^i- \Bar{\epsilon}\gamma^a\lambda\gamma_a\psi_\mu^i)\nonumber\\
					&+\frac{i}{16}(2E^{ij}+T^{ij}\cdot\gamma)(2\Bar{\psi}_\mu \epsilon \lambda_j+\Bar{\psi}_\mu\gamma_a \lambda \gamma^a \epsilon_j- \Bar{\epsilon}\gamma^a\lambda\gamma_a\psi_{\mu,j})-\frac{1}{16}(\bar{\psi}_\mu\gamma^b\lambda\gamma_a\gamma_b \bar{\lambda}\epsilon\nonumber\\
					&-\bar{\epsilon}\gamma^b\lambda \bar{\lambda}\gamma_a\gamma_b\psi_\mu)\gamma^a\lambda^i-\frac{i}{32\alpha}\bigg(-2\alpha\Bar{\epsilon}^i \gamma^{[a}\gamma_{cd}\gamma^{b]}\psi_\mu^jT^{cd}_{ij}+2\Bar{\epsilon} \psi_\mu G^{ab}-i\alpha(\Bar{\lambda}\gamma^{ab}\lambda\Bar{\epsilon}\psi_\mu\nonumber\\
					&+\Bar{\lambda}\gamma^{[a}\gamma_c\gamma^{b]} \lambda \Bar{\epsilon}\gamma^c\psi_\mu)\bigg)\gamma_{ab}\lambda^i-\frac{i}{4}E^{ij}\lambda_j\bar{\psi}_\mu\epsilon-\frac{1}{2}\bar{\psi}_\mu\gamma^a\epsilon D_a\lambda \nonumber\\
					&-\frac{i}{12} T_{\mu c}^{jk}T_{jk}^{bc}\gamma_b\epsilon^i+\frac{i}{192}\bigg(\gamma_\mu\gamma^{cd}\gamma^{ef}-\gamma^{ef}\gamma^{cd}\gamma_\mu\bigg)\epsilon^iR(M)_{cd,ef}\nonumber\\
					&-\frac{1}{24}\gamma^b\epsilon^i\bar{\lambda}R(Q)_{\mu b}+\frac{i}{24}e_{\mu,a}\gamma_b\epsilon^i\bar{\lambda}^j\bigg\{\gamma^{[a}\bigg(\frac{1}{2}D^{b]}\lambda_j-\frac{3i}{8\alpha}G^{bc}\gamma_c \lambda_j\nonumber\\
					&+\frac{i}{8}T_{jk}\cdot\gamma \gamma^{b]}\lambda^{k}\bigg)+\gamma^{ab}\bigg(\frac{1}{10\alpha}\cancel{D}\alpha\lambda_j-\frac{1}{10}\cancel{D}\lambda_j+\frac{i}{40\alpha}G\cdot \gamma\lambda_j+\frac{i}{40}T_{jk}\cdot\gamma\lambda^{k}\bigg)\bigg\}\nonumber \\
					&-\frac{i}{3}e_{\mu,a}\gamma_b\bigg(\frac{i}{48}\gamma_c G^{c[a}G^{b]d}\gamma_d\epsilon^{i}-\frac{1}{8\alpha}\gamma_cG^{c[a}\gamma\cdot T^{ij}\gamma^{b]}\epsilon_{j}-\frac{i}{4\alpha}\gamma_cG^{c[a}\lambda^i\bar{\epsilon}\gamma^{b]}\lambda\nonumber\\
					&+\frac{1}{96}\gamma\cdot T^{ij}\gamma^{[a}G^{b]c}\gamma_c\epsilon_j+\frac{1}{8}\gamma\cdot T^{ij}\gamma^{[a}\gamma\cdot T_{jk}\gamma^{b]}\epsilon^{k}-\frac{i}{2}\gamma\cdot T^{ij}\gamma^{[a}\lambda_j\bar{\epsilon}\gamma^{b]}\lambda\nonumber\\
					&-\frac{i}{48}\lambda^i \bar{\lambda}\gamma^{[a}G^{b]c}\gamma_c\epsilon+\frac{i}{4}\lambda^i\bar{\lambda}^j\gamma^{[a}\gamma\cdot T_{jk}\gamma^{b]}\epsilon^k\bigg)+\frac{i}{24}\gamma_\mu\gamma_{ef}\bigg(\frac{i}{48}\gamma_c G^{ce}G^{fd}\gamma_d\epsilon^{i}\nonumber\\
					&-\frac{1}{8\alpha}G^{ce}\gamma\cdot T^{ij}\gamma^{f}\epsilon_{j}-\frac{i}{4\alpha}\gamma_cG^{ce}\lambda^i\bar{\epsilon}\gamma^{f}\lambda+\frac{1}{96}\gamma\cdot T^{ij}\gamma^{e}G^{fc}\gamma_c\epsilon_j\nonumber\\
					&+\frac{1}{8}\gamma\cdot T^{ij}\gamma^{e}\gamma\cdot T_{jk}\gamma^{f}\epsilon^{k}-\frac{i}{2}\gamma\cdot T^{ij}\gamma^{e}\lambda_j\bar{\epsilon}\gamma^{f}\lambda-\frac{i}{48}\lambda^i \bar{\lambda}\gamma^{e}G^{fc}\gamma_c\epsilon\nonumber\\
					&+ \frac{i}{4}\lambda^i\bar{\lambda}^j\gamma^{e}\gamma\cdot T_{jk}\gamma^{f}\epsilon^k\bigg)+\frac{10i}{3}\bar{\phi}_\mu^{(i}\epsilon^{j)}\lambda_{j}+\frac{i}{8}\bar{\epsilon}\gamma^{ab}\phi_\mu\gamma_{ab}\lambda^i+\frac{1}{16}\bar{\lambda}\gamma_{ab}\phi_\mu\gamma^{ab}\epsilon^i\nonumber\\
					&-\frac{i}{4}\bar{\phi}_\mu\gamma_a\lambda\gamma^a\epsilon^i-\frac{i}{4}\bar{\epsilon}\gamma^a\phi_\mu\gamma_a\lambda^i-\frac{1}{4}\bar{\epsilon}\gamma^a\lambda\gamma_a\phi_\mu^i-\frac{10i}{3}\bar{\eta}^{(i}\psi_\mu^{j)}\lambda_{j}-\frac{i}{8}\bar{\psi}_\mu\gamma^{ab}\eta\gamma_{ab}\lambda^i\nonumber\\
					&-\frac{1}{16}\bar{\lambda}\gamma_{ab}\eta\gamma^{ab}\psi_\mu^i+\frac{i}{4}\bar{\eta}\gamma_a\lambda\gamma^a\psi_{\mu}^i+\frac{i}{4}\bar{\psi}_\mu\gamma^a\eta\gamma_a\lambda^i+\frac{1}{4}\bar{\psi}_\mu\gamma^a\lambda\gamma_a\eta^i+f_\mu^a\gamma_a\epsilon^i\nonumber\\
					\delta f_{\mu}{}^a &=\partial_\mu \Lambda^a+\Lambda f_{\mu}{}^a-\Lambda^{ab}f_{\mu b}\nonumber\\
					&-\frac{1}{24} T^{ac}_{ij}T^{ij}_{bc}\bar{\epsilon}\gamma^b\psi_\mu+\frac{1}{384}\bar{\epsilon}(\gamma^a\gamma_{cd}\gamma^{ef}-\gamma^{ef}\gamma_{cd}\gamma^a)\psi_\mu R(M)_{cd,ef}\nonumber\\
					&-\frac{i}{48}\bar{\psi}_\mu\gamma_b\epsilon\bar{\lambda}R(Q)^{ab}-\frac{1}{48}\bar{\psi}_\mu\gamma_b\epsilon\bar{\lambda}^i\bigg\{\gamma^{[a}\bigg(\frac{1}{2}D^{b]}\lambda_i-\frac{3i}{8\alpha}G^{bc}\gamma_c \lambda_i\nonumber\\
					&+\frac{i}{8}T_{ij}\cdot\gamma \gamma^{b]}\lambda_{j}\bigg)+\gamma^{ab}\bigg(\frac{1}{10\alpha}\cancel{D}\alpha\lambda_i-\frac{1}{10}\cancel{D}\lambda^i+\frac{i}{40\alpha}G\cdot \gamma\lambda^i-\frac{i}{40}T_{ij}\cdot\gamma\lambda^{j}\bigg)\bigg\}\nonumber\\
					&-\frac{i}{4}\bar{\epsilon}^i\bigg\{-\frac{i}{3}\gamma_b\bigg(\frac{i}{48}\gamma_c G^{c[a}G^{b]d}\gamma_d\psi_{\mu,i}+\frac{1}{8\alpha}\gamma_cG^{c[a}\gamma\cdot T_{ij}\gamma^{b]}\psi_{\mu}^j-\frac{i}{4\alpha}\gamma_cG^{c[a}\lambda_i\bar{\psi}_\mu\gamma^{b]}\lambda\nonumber\\
					&-\frac{1}{96}\gamma\cdot T_{ij}\gamma^{[a}G^{b]c}\gamma_c\psi_\mu^j+\frac{1}{8}\gamma\cdot T_{ij}\gamma^{[a}\gamma\cdot T^{jk}\gamma^{b]}\psi_{\mu,k}+\frac{i}{2}\gamma\cdot T_{ij}\gamma^{[a}\lambda^j\bar{\psi}_\mu\gamma^{b]}\lambda\nonumber\\
					&-\frac{i}{48}\lambda_i \bar{\lambda}\gamma^{[a}G^{b]c}\gamma_c\psi_\mu+\frac{i}{4}\lambda_i\bar{\lambda}^j\gamma^{[a}\gamma\cdot T_{jk}\gamma^{b]}\psi_\mu^k\bigg)+\frac{i}{24}\gamma^a\gamma_{ef}\bigg(\frac{i}{48}\gamma_c G^{ce}G^{fd}\gamma_d\psi_{\mu,i}\nonumber\\
					&+\frac{1}{8\alpha}\gamma_cG^{ce}\gamma\cdot T_{ij}\gamma^{f}\psi_{\mu}^j-\frac{i}{4\alpha}\gamma_cG^{ce}\lambda_i\bar{\psi}_\mu\gamma^{f}\lambda-\frac{1}{96}\gamma\cdot T_{ij}\gamma^{e}G^{fc}\gamma_c\psi_\mu^j\nonumber\\
					&+\frac{1}{8}\gamma\cdot T_{ij}\gamma^{e}\gamma\cdot T^{jk}\gamma^{f}\psi_{\mu,k}+\frac{i}{2}\gamma\cdot T_{ij}\gamma^{e}\lambda^j\bar{\psi}_\mu\gamma^{f}\lambda-\frac{i}{48}\lambda_i \bar{\lambda}\gamma^{e}G^{fc}\gamma_c\psi_\mu\nonumber\\
					&+\frac{i}{4}\lambda_i\bar{\lambda}^j\gamma^{e}\gamma\cdot T_{jk}\gamma^{f}\psi_\mu^k\bigg)\bigg\}\nonumber\\
					&+\frac{i}{4}\bar{\psi}_\mu^i\bigg\{-\frac{i}{3}\gamma_b\bigg(\frac{i}{48}\gamma_c G^{c[a}G^{b]d}\gamma_d\epsilon_{i}+\frac{1}{8\alpha}\gamma_cG^{c[a}\gamma\cdot T_{ij}\gamma^{b]}\epsilon^{j}-\frac{i}{4\alpha}\gamma_cG^{c[a}\lambda_i\bar{\epsilon}\gamma^{b]}\lambda\nonumber\\
					&-\frac{1}{96}\gamma\cdot T_{ij}\gamma^{[a}G^{b]c}\gamma_c\epsilon^j+\frac{1}{8}\gamma\cdot T_{ij}\gamma^{[a}\gamma\cdot T^{jk}\gamma^{b]}\epsilon_{k}+\frac{i}{2}\gamma\cdot T_{ij}\gamma^{[a}\lambda^j\bar{\epsilon}\gamma^{b]}\lambda\nonumber\\
					&-\frac{i}{48}\lambda_i \bar{\lambda}\gamma^{[a}G^{b]c}\gamma_c\epsilon+\frac{i}{4}\lambda_i\bar{\lambda}^j\gamma^{[a}\gamma\cdot T_{jk}\gamma^{b]}\epsilon^k\bigg)+\frac{i}{24}\gamma^a\gamma_{ef}\bigg(\frac{i}{48}\gamma_c G^{ce}G^{fd}\gamma_d\epsilon_{i}\nonumber\\
					&+\frac{1}{8\alpha}\gamma_cG^{ce}\gamma\cdot T_{ij}\gamma^{f}\epsilon^{j}-\frac{i}{4\alpha}\gamma_cG^{ce}\lambda_i\bar{\epsilon}\gamma^{f}\lambda-\frac{1}{96}\gamma\cdot T_{ij}\gamma^{e}G^{fc}\gamma_c\epsilon^j\nonumber\\
					&+\frac{1}{8}\gamma\cdot T_{ij}\gamma^{e}\gamma\cdot T^{jk}\gamma^{f}\epsilon_{k}+\frac{i}{2}\gamma\cdot T_{ij}\gamma^{e}\lambda^j\bar{\epsilon}\gamma^{f}\lambda-\frac{i}{48}\lambda_i \bar{\lambda}\gamma^{e}G^{fc}\gamma_c\epsilon\nonumber\\
					&+\frac{i}{4}\lambda_i\bar{\lambda}^j\gamma^{e}\gamma\cdot T_{jk}\gamma^{f}\epsilon_1^k\bigg)\bigg\}-\frac{1}{48}e_\mu^a T^{cd}_{ij}\biggl\{-\frac{i}{8}\bar{\epsilon}^{[i}\gamma^{ef}\gamma_{cd}R(Q)_{ef}^{j]}+\frac{i}{16}\bar{\epsilon}^k\gamma_{cd}\chi_k{}^{ij}\nonumber\\
					&-\frac{i}{8\alpha}\bar{\epsilon}^{[i}\gamma \cdot G\gamma_{cd}\lambda^{j]}-\frac{i}{4}\bar{\epsilon}\gamma_{cdef}\lambda T^{ef,ij}-\frac{1}{4}\bar{\epsilon}^{[i}\gamma^e\gamma_{cd}\bigg(-\frac{1}{\alpha}D_e\alpha\lambda^{j]}+D_e\lambda^{j]}\nonumber\\
					&+\frac{i}{4\alpha}G_{ef}\gamma^f\lambda^{j]}-\frac{i}{4}\gamma\cdot T^{j]k} \gamma_e\lambda_k\bigg)-\frac{1}{20}\bar{\epsilon}^{[i}\gamma_{cd}\bigg(\frac{1}{\alpha}\cancel{D}\alpha\lambda^{j]}-\cancel{D}\lambda^{j]}-\frac{i}{4\alpha}\gamma\cdot G \lambda^{j]}\nonumber\\
					&+\frac{i}{4}\gamma\cdot T^{j]k}\lambda_k\bigg)-\Omega\text{trace}\biggr\}+\frac{1}{12}T^{ab}_{ij}\biggl\{-\frac{i}{8}\bar{\epsilon}^{[i}\gamma^{cd}\gamma_{\mu b}R(Q)_{cd}^{j]}+\frac{i}{16}\bar{\epsilon}^k\gamma_{\mu b}\chi_k{}^{ij}\nonumber\\
					&-\frac{i}{8\alpha}\bar{\epsilon}^{[i}\gamma \cdot G\gamma_{\mu b}\lambda^{j]}-\frac{i}{4}\bar{\epsilon}\gamma_{\mu bcd}\lambda T^{cd,ij}-\frac{1}{4}\bar{\epsilon}^{[i}\gamma^c\gamma_{\mu b}\bigg(-\frac{1}{\alpha}D_c\alpha\lambda^{j]}+D_c\lambda^{j]}\nonumber\\
					&+\frac{i}{4\alpha}G_{cd}\gamma^d\lambda^{j]}-\frac{i}{4}\gamma\cdot T^{j]k} \gamma_c\lambda_k\bigg)-\frac{1}{20}\bar{\epsilon}^{[i}\gamma_{\mu b}\bigg(\frac{1}{\alpha}\cancel{D}\alpha\lambda^{j]}-\cancel{D}\lambda^{j]}-\frac{i}{4\alpha}\gamma\cdot G \lambda^{j]}\nonumber\\
					&+\frac{i}{4}\gamma\cdot T^{j]k}\lambda_k\bigg)-\Omega\text{trace}\biggr\}+\frac{1}{12}T^{ij}_{\mu b}\biggl\{-\frac{i}{8}\bar{\epsilon}^{[i}\gamma^{cd}\gamma^{ab}R(Q)_{cd}^{j]}+\frac{i}{16}\bar{\epsilon}^k\gamma^{ab}\chi_k{}^{ij}\nonumber\\
					&-\frac{i}{8\alpha}\bar{\epsilon}^{[i}\gamma \cdot G\gamma^{ab}\lambda^{j]}-\frac{i}{4}\bar{\epsilon}\gamma^{ab}{}_{cd}\lambda T^{cd,ij}-\frac{1}{4}\bar{\epsilon}^{[i}\gamma^c\gamma^{ab}\bigg(-\frac{1}{\alpha}D_c\alpha\lambda^{j]}+D_c\lambda^{j]}\nonumber\\
					&+\frac{i}{4\alpha}G_{cd}\gamma^d\lambda^{j]}-\frac{i}{4}\gamma\cdot T^{j]k} \gamma_c\lambda_k\bigg)-\frac{1}{20}\bar{\epsilon}^{[i}\gamma^{ab}\bigg(\frac{1}{\alpha}\cancel{D}\alpha\lambda^{j]}-\cancel{D}\lambda^{j]}-\frac{i}{4\alpha}\gamma\cdot G \lambda^{j]}\nonumber\\
					&+\frac{i}{4}\gamma\cdot T^{j]k}\lambda_k\bigg)-\Omega\text{trace}\biggr\}-\frac{1}{6}\bigg\{-\frac{i}{4}\bar{\epsilon}^i\gamma^{[a}\gamma_{cd}\gamma^{b]}R(Q)_{\mu b}^jT_{ij}^{cd}\nonumber\\
					&-\frac{i}{2}\bar{\epsilon}\gamma^{ab}R(S)_{\mu b}+\frac{i}{4\alpha}\bar{\epsilon}R(Q)_{\mu b}G^{ab}+\frac{1}{8}\bigg(\bar{\lambda}\gamma_{ab}\lambda\bar{\epsilon}R(Q)_{\mu b}\nonumber\\
					&+\bar{\lambda}\gamma_{[a}\gamma_c\gamma_{b]}\lambda\bar{\epsilon}\gamma^cR(Q)_{\mu b}\bigg)\bigg\}-\frac{1}{6}e^\nu{}_b\biggl\{-\frac{1}{2}\bar{R}(Q)^{ab}\gamma_{[\nu}\left(\frac{i}{4}G_{\mu]c}\gamma^c\epsilon+\frac{1}{2}\lambda\bar{\epsilon}\gamma_{\mu]}\lambda\right)\nonumber\\
					&-\frac{i}{8}\bar{R}(Q)^{ab,i}\gamma_{[\nu}\gamma^{cd}\gamma_{\mu]}\epsilon^j T^{cd}_{ij}+\bar{R}(Q)_{[\nu}{}^{[a}\gamma^{b]}\left(\frac{i}{4}G_{\mu]c}\gamma^c\epsilon+\frac{1}{2}\lambda\bar{\epsilon}\gamma_{\mu]}\lambda \right)\nonumber\\
					&+\frac{i}{4}\bar{R}(Q)^i{}_{[\nu}{}^{[a}\gamma^{b]}T_{cd,ij}\gamma^{cd}\gamma_{\mu]}\epsilon^j\biggr\}+\frac{1}{48}e_\mu{}^a\biggl\{ -\frac{1}{2}\bar{R}(Q)^{cd}\gamma_{d}\bigg(\frac{i}{4}G_{ce}\gamma^e\epsilon+\frac{1}{2}\lambda\bar{\epsilon}\gamma_c\lambda\bigg) \nonumber\\
					&-\frac{i}{8}\bar{R}(Q)^{cd,i}\gamma_d\gamma^{ef}\gamma_c\epsilon^j T_{ef,ij}+\bar{R}(Q)_{d}{}^{[c}\gamma^{d]}\bigg(\frac{i}{4}G_{ce}\gamma^e\epsilon-\frac{1}{2}\lambda\bar{\epsilon}\gamma_c\lambda\bigg)\nonumber\\
					&+\frac{i}{4}\bar{R}(Q)^i{}_{d}{}^{[c}\gamma^{d]}T_{ef,ij}\gamma^{ef}\gamma_c\epsilon^j-\frac{i}{4}\bar{\epsilon}^i\gamma^{e}\gamma_{cd}\gamma^{f}R(Q)_{ef}^jT_{ij}^{cd}-\frac{i}{2}\bar{\epsilon}\gamma^{cd}R(S)_{cd}\nonumber\\
					&+\frac{i}{4\alpha}\bar{\epsilon}R(Q)_{cd}G^{cd}+\frac{1}{8}\bigg(\bar{\lambda}\gamma^{cd}\lambda\bar{\epsilon}R(Q)_{cd}+\bar{\lambda}\gamma^{[c}\gamma_c\gamma^{d]}\lambda\bar{\epsilon}\gamma^cR(Q)_{cd}\bigg)\biggr\}-\frac{i}{12}\bar{\eta}R(Q)_{\mu}{}^{a}\nonumber\\
					&+\frac{1}{12}\bar{\eta}^ie_{\mu,b}\bigg\{\gamma^{[a}(-\frac{1}{2\alpha}D^{b]}\alpha \lambda_i+\frac{1}{2}D^{b]}\lambda_i-\frac{3i}{8\alpha}G^{b]c}\gamma_c\lambda_i+\frac{i}{8}T_{ij}\cdot\gamma \gamma^{b]}\lambda^{j})+\gamma^{ab}(\frac{1}{10\alpha}\cancel{D}\alpha\lambda_i\nonumber\\
					&-\frac{1}{10}\cancel{D}\lambda_i+\frac{i}{40\alpha}G\cdot \gamma\lambda_i-\frac{i}{40}T_{ij}\cdot\gamma\lambda^{j})\bigg\}+\frac{i}{4}\bar{\phi}_\mu^i\bigg(\frac{i}{4\alpha}G^{ab}\gamma_b\epsilon_i+\frac{i}{4}T_{cd,ij}\gamma^{cd}\gamma^a\epsilon^j\nonumber\\
					&+\frac{1}{2}\bar{\epsilon}\gamma^a\lambda\lambda_i\bigg)-\frac{i}{4}\bar{\eta}^i\bigg(\frac{i}{4\alpha}G^{ab}\gamma_b\psi_{\mu,i}+\frac{i}{4}T_{cd,ij}\gamma^{cd}\gamma^a\psi_\mu^j+\frac{1}{2}\bar{\psi}_\mu\gamma^a\lambda\lambda_i\bigg)+\frac{1}{2}\bar{\eta}\gamma^a\phi_\mu\nonumber\\
				\end{align}	
			}

			The corresponding curvatures are given below:
			{\allowdisplaybreaks
				\begin{align}
					R(S)_{\mu \nu}^i&=2\partial_{[\mu}\phi_{\nu]}^i-b_{[\mu} \phi_{\nu ]}^{i}+V_{[\mu}^{ij}\phi_{\nu],j}+\frac{1}{2}\omega_{[\mu}^{ab}\gamma_{ab}\phi^i_{\nu]}+2\gamma_af^a_{[\mu}\psi_{\nu]}^i\nonumber\\
				&-2\biggl[\gamma^{ab}\psi_{[\nu}^i\bigg(-\frac{i}{16}\bar{\psi}_{\mu]} R(Q)_{ab}+\frac{1}{8}\bar{\psi}_{\mu]}\gamma_a D_b\lambda\nonumber \\
				&-\frac{1}{8\alpha}D_a\alpha \bar{\psi}_{\mu]}\gamma_b\lambda+\frac{i}{32\alpha}\bar{\lambda}\gamma_a\gamma_c G_{bc}\psi_{\mu]}+\frac{i}{32}\bar{\lambda}^j\gamma_a\gamma\cdot T_{jk}\gamma_b\psi_{\mu]}^k\bigg)+\gamma^a\psi_{[\nu}^i\bigg(\frac{1}{4}\bar{\psi}_{\mu]} D_a\lambda\nonumber\\
				&+\frac{i}{16\alpha}\bar{\lambda}G_{ac}\gamma^c\psi_{\mu]}+\frac{i}{16}\bar{\lambda}^j\gamma\cdot T_{jk}\gamma_a\psi_{\mu]}^k\bigg)+\frac{1}{4}\gamma^{ab}\psi_{j,[\nu}\biggl\{\frac{i}{8\alpha}\bar{\psi}_{\mu]}^{[i}\gamma \cdot G\gamma_{ab}\lambda^{j]}+\frac{1}{4}\bar{\psi}_{\mu]}^{[i}\gamma^c\gamma_{ab}\nonumber\\
				&\bigg(-\frac{1}{\alpha}D_c\alpha\lambda^{j]}+D_c\lambda^{j]}+\frac{i}{4\alpha}G_{cd}\gamma^d\lambda^{j]}-\frac{i}{4}T^{j]k}\cdot\gamma \gamma_c\lambda_k\bigg)+\frac{1}{20}\bar{\psi}_{\mu]}^{[i}\gamma_{ab}\bigg(\frac{1}{\alpha}\cancel{D}\alpha\lambda^{j]}\nonumber\\
				&v-\frac{i}{4\alpha}G\cdot\gamma \lambda^{j]}+\frac{i}{4}T^{j]k}\cdot\gamma\lambda_k\bigg)-\frac{i}{16}\bar{\psi}_{\mu]}^k\gamma_{ab}\chi_k^{ij}-\frac{i}{8}\bar{\psi}_{\mu]}^{[i}\gamma^{cd}\gamma_{ab}R(Q)_{cd}^{j]}\nonumber\\
				&+\frac{i}{4}\bar{\psi}_{\mu]}\gamma_{abcd}\lambda T^{cd,ij}-\Omega\text{trace}\biggr\}-\frac{i}{4}\bar{\psi}_{[\nu}\gamma^a\lambda\gamma_a\bigg(-\frac{1}{8\alpha} G \cdot \gamma \psi_{\mu]}^i 
				+ \frac{i}{2\alpha} \cancel{D}\alpha\psi_{\mu]}^i-\frac{1}{2}E^i_{~j}\psi_{\mu]}^j 
				\nonumber\\
				&+\frac{1}{4}\gamma \cdot T^{ij}\psi_{\mu],j} 
				+\frac{i}{4}\bar{\psi}_{\mu]}\gamma ^a \lambda \gamma _a \lambda^i+\frac{i}{2}\bar{\psi}_{\mu]}\lambda\lambda^i\bigg)+\frac{1}{2}\psi_{j,[\nu}\bigg(-\frac{4}{5\alpha}\bar{\psi}_{\mu]}^{(i} \lambda^{j)} \cancel{D}\alpha 
				\nonumber\\
				&	+\frac{4}{5}\bar{\psi}_{\mu]}^{(i}\cancel{D}\lambda^{j)}-i\bar{\psi}_{\mu]}^{(i}T^{j)k} \cdot \gamma \lambda_k
				-\frac{i}{4}\bar{\psi}_{\mu],k}\chi^{(i,j)k}
				-\frac{i}{2}E^{ij}\bar{\psi}_{\mu]}\lambda\bigg)-\frac{i}{2}\bar{\psi}_{[\nu}\lambda\bigg(-\frac{1}{8\alpha}\gamma \cdot G \psi_{\mu]}^i 
				\nonumber\\
				&	+ \frac{i}{2\alpha} \cancel{D} \alpha \psi_{\mu]}^i 
				-\frac{1}{2}E^i_{~j}\psi_{\mu]}^j+\frac{1}{4}\gamma\cdot T^{ij}\psi_{\mu],j}+\frac{i}{2}\bar{\psi}_{\mu]}\lambda\lambda^i 
				+\frac{i}{2}\bar{\psi}_{\mu]}\gamma ^a \lambda \gamma _a \lambda^i\bigg)\nonumber\\
				&-\gamma_a\lambda^i\bigg(-\frac{1}{16}\bar{\psi}_{\mu]}\gamma_b\lambda\bar{\epsilon}\gamma^a\gamma^b\lambda-\frac{1}{8}\bar{\psi}_{\mu]}\lambda\bar{\epsilon}\gamma^a\lambda\bigg)\biggr]+i\lambda\bigg(\frac{i}{2\alpha}\bar{\psi}_{[\nu}\cancel{D}\alpha\psi_{\mu]}\nonumber\\
						&-\frac{1}{4}\bar{\psi}_{[\nu}^j\gamma\cdot T_{jk}\psi_{\mu]}^k\bigg)-\gamma_b\lambda^i\bigg(\frac{i}{8\alpha}\bar{\psi}_{[\nu}\gamma_a\psi_{\mu]}G^{ab}+\frac{1}{4\alpha}\bar{\psi}_{[\nu}D^b\alpha\psi_{\mu]}+\frac{i}{4}\bar{\psi}_{[\nu}^j\gamma_a\psi_{\mu]}^k T^{ab}_{jk}\bigg)\nonumber\\
				&-\frac{i}{32\alpha}(G\cdot\gamma-4i\cancel{D}\alpha)(2\Bar{\psi}_{[\mu} \psi_{\nu]} \lambda^i+\Bar{\psi}_{[\mu}\gamma_a \lambda \gamma^a \psi_{\nu]}^i- \Bar{\psi}_{[\nu}\gamma^a\lambda\gamma_a\psi_{\mu]}^i)+\frac{i}{16}(2E^{ij}\nonumber\\
				&+T^{ij}\cdot\gamma)(2\Bar{\psi}_{[\mu} \psi_{\nu]} \lambda_j+\Bar{\psi}_{[\mu}\gamma_a \lambda \gamma^a \psi_{\nu],j}- \Bar{\psi}_{[\nu}\gamma^a\lambda\gamma_a\psi_{\mu],j})-\frac{1}{16}(\bar{\psi}_{[\mu}\gamma^b\lambda\gamma_a\gamma_b \bar{\lambda}\psi_{\nu]}\nonumber\\
				&-\bar{\psi}_{[\nu}\gamma^b\lambda \bar{\lambda}\gamma_a\gamma_b\psi_{\mu]})\gamma^a\lambda^i-\frac{i}{32\alpha}\bigg(-2\alpha\Bar{\psi}_{[\nu}^i \gamma^{[a}\gamma_{cd}\gamma^{b]}\psi_{\mu]}^jT^{cd}_{ij}+2\Bar{\psi}_{[\nu} \psi_{\mu]} G^{ab}\nonumber\\
				&-i\alpha(\Bar{\lambda}\gamma^{ab}\lambda\Bar{\psi}_{[\nu}\psi_{\mu]}+\Bar{\lambda}\gamma^{[a}\gamma_c\gamma^{b]} \lambda \Bar{\psi}_{[\nu}\gamma^c\psi_{\mu]})\bigg)\gamma_{ab}\lambda^i-\frac{i}{4}E^{ij}\lambda_j\bar{\psi}_{[\mu}\psi_{\nu]}-\frac{1}{2}\bar{\psi}_{[\mu}\gamma^a\psi_{\nu]} D_a\lambda \nonumber\\
				&+2\bigg[\frac{i}{12} T_{c[\mu}^{jk}T_{jk}^{bc}\gamma_b\psi_{\nu]}^i+\frac{i}{192}\bigg(\gamma_{[\mu}\gamma^{cd}\gamma^{ef}-\gamma^{ef}\gamma^{cd}\gamma_{[\mu}\bigg)\psi_{\nu]}^iR(M)_{cd,ef}\nonumber\\
				&-\frac{1}{24}\gamma^b\psi_{[\nu}^i\bar{\lambda}R(Q)_{\mu] b}+\frac{i}{24}e_{[\mu,a}\gamma_b\psi_{\nu]}^i\bar{\lambda}^j\bigg\{\gamma^{[a}\bigg(\frac{1}{2}D^{b]}\lambda_j-\frac{3i}{8\alpha}G^{bc}\gamma_c \lambda_j\nonumber\\
				&+\frac{i}{8}T_{jk}\cdot\gamma \gamma^{b]}\lambda^{k}\bigg)+\gamma^{ab}\bigg(\frac{1}{10\alpha}\cancel{D}\alpha\lambda_j-\frac{1}{10}\cancel{D}\lambda_j+\frac{i}{40\alpha}G\cdot \gamma\lambda_j+\frac{i}{40}T_{jk}\cdot\gamma\lambda^{k}\bigg)\bigg\}\nonumber \\
				&-\frac{i}{3}e_{[\mu,a}\gamma_b\bigg(\frac{i}{48}\gamma_c G^{c[a}G^{b]d}\gamma_d\psi_{\nu]}^{i}-\frac{1}{8\alpha}\gamma_cG^{c[a}\gamma\cdot T^{ij}\gamma^{b]}\psi_{\nu],j}-\frac{i}{4\alpha}\gamma_cG^{c[a}\lambda^i\bar{\psi}_{\nu]}\gamma^{b]}\lambda\nonumber\\
				&+\frac{1}{96}\gamma\cdot T^{ij}\gamma^{[a}G^{b]c}\gamma_c\psi_{\nu],j}+\frac{1}{8}\gamma\cdot T^{ij}\gamma^{[a}\gamma\cdot T_{jk}\gamma^{b]}\psi_{\nu]}^{1,k}-\frac{i}{2}\gamma\cdot T^{ij}\gamma^{[a}\lambda_j\bar{\psi}_{\nu]}\gamma^{b]}\lambda\nonumber\\
				&-\frac{i}{48}\lambda^i \bar{\lambda}\gamma^{[a}G^{b]c}\gamma_c\psi_{\nu]}+\frac{i}{4}\lambda^i\bar{\lambda}^j\gamma^{[a}\gamma\cdot T_{jk}\gamma^{b]}\psi_{\nu]}^k\bigg)+\frac{i}{24}\gamma_{[\mu}\gamma_{ef}\bigg(\frac{i}{48}\gamma_c G^{ce}G^{fd}\gamma_d\psi_{\nu]}^{i}\nonumber\\
				&-\frac{1}{8\alpha}G^{ce}\gamma\cdot T^{ij}\gamma^{f}\psi_{\nu],j}-\frac{i}{4\alpha}\gamma_cG^{ce}\lambda^i\bar{\psi}_{\nu]}\gamma^{f}\lambda+\frac{1}{96}\gamma\cdot T^{ij}\gamma^{e}G^{fc}\gamma_c\psi_{\nu],j}\nonumber\\
				&+\frac{1}{8}\gamma\cdot T^{ij}\gamma^{e}\gamma\cdot T_{jk}\gamma^{f}\psi_{\nu]}^{k}-\frac{i}{2}\gamma\cdot T^{ij}\gamma^{e}\lambda_j\bar{\psi}_{\nu]}\gamma^{f}\lambda-\frac{i}{48}\lambda^i \bar{\lambda}\gamma^{e}G^{fc}\gamma_c\psi_{\nu]}\nonumber\\
				&+ \frac{i}{4}\lambda^i\bar{\lambda}^j\gamma^{e}\gamma\cdot T_{jk}\gamma^{f}\psi_{\nu]}^k\bigg)\bigg]\nonumber\\
					&+\frac{20i}{3}\bar{\phi}_{[\mu}^{(i}\psi_{\nu]}^{j)}\lambda_{j}+\frac{i}{4}\bar{\psi}_{[\nu}\gamma^{ab}\phi_{\mu]}\gamma_{ab}\lambda^i+\frac{1}{8}\bar{\lambda}\gamma_{ab}\phi_{[\mu}\gamma^{ab}\psi_{\nu]}^i\nonumber\\
					&-\frac{i}{2}\bar{\phi}_{[\mu}\gamma_a\lambda\gamma^a\psi_{\nu]}^i-\frac{i}{2}\bar{\psi}_{[\nu}\gamma^a\phi_{\mu]}\gamma_a\lambda^i-\frac{1}{2}\bar{\psi}_{[\nu}\gamma^a\lambda\gamma_a\phi_{\mu]}^i\nonumber\\
					R(K)_{\mu\nu}{}^a&=2\partial_{[\mu} f_{\nu]}{}^a-2b_{[\mu}f_{\nu]}{}^a-2f_{b[\nu }\omega_{\mu]}{}^{ab}-\frac{1}{24} T^{ac}_{ij}T^{ij}_{bc}\bar{\psi}_{[\nu}\gamma^b\psi_{\mu]}\nonumber\\
					&+\frac{1}{384}\bar{\psi}_{[\nu}(\gamma^a\gamma_{cd}\gamma^{ef}-\gamma^{ef}\gamma_{cd}\gamma^a)\psi_{\mu]} R(M)_{cd,ef}\nonumber\\
					&-\frac{i}{48}\bar{\psi}_{[\mu}\gamma_b\psi_{\nu]}\bar{\lambda}R(Q)^{ab}-\frac{1}{48}\bar{\psi}_{[\mu}\gamma_b\psi_{\nu]}\bar{\lambda}^i\bigg\{\gamma^{[a}\bigg(\frac{1}{2}D^{b]}\lambda_i-\frac{3i}{8\alpha}G^{bc}\gamma_c \lambda_i\nonumber\\
					&+\frac{i}{8}T_{ij}\cdot\gamma \gamma^{b]}\lambda_{j}\bigg)+\gamma^{ab}\bigg(\frac{1}{10\alpha}\cancel{D}\alpha\lambda_i-\frac{1}{10}\cancel{D}\lambda_i+\frac{i}{40\alpha}G\cdot \gamma\lambda_i-\frac{i}{40}T_{ij}\cdot\gamma\lambda^{j}\bigg)\bigg\}\nonumber\\
					&+\frac{i}{2}\bar{\psi}_{[\nu}^i\bigg\{-\frac{i}{3}\gamma_b\bigg(\frac{i}{48}\gamma_c G^{c[a}G^{b]d}\gamma_d\psi_{\mu],i}-\frac{1}{8\alpha}G^{c[a}\gamma\cdot T^{ij}\gamma^{b]}\psi_{\mu],j}-\frac{i}{4\alpha}\gamma_cG^{c[a}\lambda_i\bar{\psi}_{\mu]}\gamma^{b]}\lambda\nonumber\\
					&-\frac{1}{96}\gamma\cdot T_{ij}\gamma^{[a}G^{b]c}\gamma_c\psi_{\mu]}^j+\frac{1}{8}\gamma\cdot T_{ij}\gamma^{[a}\gamma\cdot T^{jk}\gamma^{b]}\psi_{\mu],k}+\frac{i}{2}\gamma\cdot T_{ij}\gamma^{[a}\lambda^j\bar{\psi}_{\mu]}\gamma^{b]}\lambda\nonumber\\
					&-\frac{i}{48}\lambda_i \bar{\lambda}\gamma^{[a}G^{b]c}\gamma_c\psi_{\mu]}+\frac{i}{4}\lambda_i\bar{\lambda}^j\gamma^{[a}\gamma\cdot T_{jk}\gamma^{b]}\psi_{\mu]}^k\bigg)+\frac{i}{24}\gamma^a\gamma_{ef}\bigg(\frac{i}{48}\gamma_c G^{ce}G^{fd}\gamma_d\psi_{\mu],i}\nonumber\\
					&+\frac{1}{8\alpha}G^{ce}\gamma\cdot T_{ij}\gamma^{f}\psi_{\mu]}^{j}-\frac{i}{4\alpha}\gamma_cG^{ce}\lambda_i\bar{\psi}_{\mu]}\gamma^{f}\lambda-\frac{1}{96}\gamma\cdot T_{ij}\gamma^{e}G^{fc}\gamma_c\psi_{\mu]}^j\nonumber\\
					&+\frac{1}{8}\gamma\cdot T_{ij}\gamma^{e}\gamma\cdot T^{jk}\gamma^{f}\psi_{\mu],k}+\frac{i}{2}\gamma\cdot T_{ij}\gamma^{e}\lambda^j\bar{\psi}_{\mu]}\gamma^{f}\lambda-\frac{i}{48}\lambda_i \bar{\lambda}\gamma^{e}G^{fc}\gamma_c\psi_{\mu]}\nonumber\\
					&+\frac{i}{4}\lambda_i\bar{\lambda}^j\gamma^{e}\gamma\cdot T_{jk}\gamma^{f}\psi_{\mu]}^k\bigg)\bigg\}-\frac{1}{3}\bigg\{-\frac{i}{4}\bar{\psi}_{[\nu}^i\gamma_{[a}\gamma_{cd}\gamma_{b]}R(Q)_{\mu] b}^jT_{ij}^{cd}-\frac{i}{2}\bar{\psi}_{[\nu}\gamma_{ab}R(S)_{\mu] b}\nonumber\\
					&+\frac{i}{4\alpha}\bar{\psi}_{[\nu}R(Q)_{\mu] b}G_{ab}+\frac{1}{8}\bigg(\bar{\lambda}\gamma_{ab}\lambda\bar{\psi}_{[\nu}R(Q)_{\mu] b}+\bar{\lambda}\gamma_{[a}\gamma_c\gamma_{b]}\lambda\bar{\psi}_{[\nu}\gamma^cR(Q)_{\mu] b}\bigg)\bigg\}\nonumber\\
					&+\frac{1}{3}e^\rho{}_b\biggl\{\frac{1}{2}\bar{R}(Q)^{ab}\gamma_{[\rho}\bigg(\frac{i}{4}G_{\mu]c}\gamma^c\psi_\nu+\frac{1}{2}\lambda\bar{\psi}_\nu\gamma_{\mu]}\lambda-\frac{i}{4}G_{\nu c}\gamma^c\psi_{\mu]}\nonumber\\
					&-\frac{1}{2}\lambda\bar{\psi}_{\mu]}\gamma_{\nu}\lambda\bigg)-\frac{i}{8}\bar{R}(Q)^{ab,i}\gamma_{[\rho}\gamma^{cd}\gamma_{\nu}\psi_{\mu]}^j T^{cd}_{ij}+\frac{i}{8}\bar{R}(Q)^{ab,i}\gamma_{[\rho}\gamma^{cd}\gamma_{\mu]}\psi_\nu^j T^{cd}_{ij}\nonumber\\
					&-\bar{R}(Q)_{[\rho}{}^{[a}\gamma^{b]}\left(\frac{i}{4}G_{\mu]c}\gamma^c\psi_\nu+\frac{1}{2}\lambda\bar{\psi}_\nu\gamma_{\mu]}\lambda-\frac{i}{4}G_{\nu c}\gamma^c\psi_{\mu]}-\frac{1}{2}\lambda\bar{\psi}_{\mu]}\gamma_{\nu}\lambda\right)\nonumber\\
					&-\frac{i}{4}\bar{R}(Q)^i{}_{[\rho}{}^{[a}\gamma^{b]}T_{cd,ij}\gamma^{cd}\gamma_{\mu]}\psi_\nu^j+\frac{i}{4}\bar{R}(Q)^i{}_{[\rho}{}^{[a}\gamma^{b]}T_{cd,ij}\gamma^{cd}\gamma_{\nu}\psi_{\mu]}^j\biggr\}\nonumber\\
					&-\frac{1}{24}e_{[\mu}{}^a\biggl\{ \frac{1}{2}\bar{R}(Q)^{cd}\gamma_{d}\bigg(\frac{i}{4}G_{ce}\gamma^e\psi_{\nu]}+\frac{1}{2}\lambda\bar{\psi}_{\nu]}\gamma_c\lambda\bigg) \nonumber\\
					&+\frac{i}{8}\bar{R}(Q)^{cd,i}\gamma_d\gamma^{ef}\gamma_c\psi_{\nu]}^j T_{ef,ij}-\bar{R}(Q)_{d}{}^{[c}\gamma^{d]}\bigg(\frac{i}{4}G_{ce}\gamma^e\psi_{\nu]}+\frac{1}{2}\lambda\bar{\psi}_{\nu]}\gamma_c\lambda\bigg)\nonumber\\
					&-\frac{i}{4}\bar{R}(Q)^i{}_{d}{}^{[c}\gamma^{d]}T_{ef,ij}\gamma^{ef}\gamma_c\psi_{\nu]}^j+\frac{i}{4}\bar{\psi}_{\nu]}^i\gamma^{e}\gamma_{cd}\gamma^{f}R(Q)_{ef}^jT_{ij}^{cd}+\frac{i}{2}\bar{\psi}_{\nu]}\gamma^{cd}R(S)_{cd}\nonumber\\
					&-\frac{i}{4\alpha}\bar{\psi}_{\nu]}R(Q)_{cd}G^{cd}-\frac{1}{8}\bigg(\bar{\lambda}\gamma^{cd}\lambda\bar{\psi}_{\nu]}R(Q)_{cd}+\bar{\lambda}\gamma^{[c}\gamma_c\gamma^{d]}\lambda\bar{\psi}_{\nu]}\gamma^cR(Q)_{cd}\bigg)\biggr\}\nonumber\\
					&+2\bigg[-\frac{1}{48}e_{[\mu}^a T^{cd}_{ij}\biggl\{-\frac{i}{8}\bar{\psi}_{\nu]}^{[i}\gamma^{ef}\gamma_{cd}R(Q)_{ef}^{j]}+\frac{i}{16}\bar{\psi}_{\nu]}^k\gamma_{cd}\chi_k{}^{ij}\nonumber\\
					&-\frac{i}{8\alpha}\bar{\psi}_{\nu]}^{[i}\gamma \cdot G\gamma_{cd}\lambda^{j]}-\frac{i}{4}\bar{\psi}_{\nu]}\gamma_{cdef}\lambda T^{ef,ij}-\frac{1}{4}\bar{\psi}_{\nu]}^{[i}\gamma^e\gamma_{cd}\bigg(-\frac{1}{\alpha}D_e\alpha\lambda^{j]}+D_e\lambda^{j]}\nonumber\\
					&+\frac{i}{4\alpha}G_{ef}\gamma^f\lambda^{j]}-\frac{i}{4}\gamma\cdot T^{j]k} \gamma_e\lambda_k\bigg)-\frac{1}{20}\bar{\psi}_{\nu]}^{[i}\gamma_{cd}\bigg(\frac{1}{\alpha}\cancel{D}\alpha\lambda^{j]}-\cancel{D}\lambda^{j]}-\frac{i}{4\alpha}\gamma\cdot G \lambda^{j]}\nonumber\\
					&+\frac{i}{4}\gamma\cdot T^{j]k}\lambda_k\bigg)-\Omega\text{trace}\biggr\}+\frac{1}{12}T^{ab}_{ij}\biggl\{-\frac{i}{8}\bar{\psi}_{[\nu}^{[i}\gamma^{cd}\gamma_{\mu] b}R(Q)_{cd}^{j]}+\frac{i}{16}\bar{\psi}_{[\nu}^k\gamma_{\mu] b}\chi_k{}^{ij}\nonumber\\
					&-\frac{i}{8\alpha}\bar{\psi}_{[\nu}^{[i}\gamma \cdot G\gamma_{\mu] b}\lambda^{j]}-\frac{i}{4}\bar{\epsilon}\gamma_{\mu ]bcd}\lambda T^{cd,ij}-\frac{1}{4}\bar{\psi}_{[\nu}^{[i}\gamma^c\gamma_{\mu] b}\bigg(-\frac{1}{\alpha}D_c\alpha\lambda^{j]}+D_c\lambda^{j]}\nonumber\\
					&+\frac{i}{4\alpha}G_{cd}\gamma^d\lambda^{j]}-\frac{i}{4}\gamma\cdot T^{j]k} \gamma_c\lambda_k\bigg)-\frac{1}{20}\bar{\psi}_{[\nu}^{[i}\gamma_{\mu] b}\bigg(\frac{1}{\alpha}\cancel{D}\alpha\lambda^{j]}-\cancel{D}\lambda^{j]}-\frac{i}{4\alpha}\gamma\cdot G \lambda^{j]}\nonumber\\
					&+\frac{i}{4}\gamma\cdot T^{j]k}\lambda_k\bigg)-\Omega\text{trace}\biggr\}-\frac{1}{12}T^{ij}_{b[\mu }\biggl\{-\frac{i}{8}\bar{\psi}_{\nu]}^{[i}\gamma^{cd}\gamma^{ab}R(Q)_{cd}^{j]}+\frac{i}{16}\bar{\psi}_{\nu]}^k\gamma^{ab}\chi_k{}^{ij}\nonumber\\
					&-\frac{i}{8\alpha}\bar{\psi}_{\nu]}^{[i}\gamma \cdot G\gamma^{ab}\lambda^{j]}-\frac{i}{4}\bar{\psi}_{\nu]}\gamma^{ab}{}_{cd}\lambda T^{cd,ij}-\frac{1}{4}\bar{\psi}_{\nu]}^{[i}\gamma^c\gamma^{ab}\bigg(-\frac{1}{\alpha}D_c\alpha\lambda^{j]}+D_c\lambda^{j]}\nonumber\\
					&+\frac{i}{4\alpha}G_{cd}\gamma^d\lambda^{j]}-\frac{i}{4}\gamma\cdot T^{j]k} \gamma_c\lambda_k\bigg)-\frac{1}{20}\bar{\psi}_{\nu]}^{[i}\gamma^{ab}\bigg(\frac{1}{\alpha}\cancel{D}\alpha\lambda^{j]}-\cancel{D}\lambda^{j]}-\frac{i}{4\alpha}\gamma\cdot G \lambda^{j]}\nonumber\\
					&+\frac{i}{4}\gamma\cdot T^{j]k}\lambda_k\bigg)-\Omega\text{trace}\biggr\}\bigg]\nonumber\\
					&-\frac{i}{6}\bar{\phi}_{[\nu}R(Q)_{\mu]}{}^{a}+\frac{1}{6}\bar{\phi}_{[\nu}^ie_{\mu],b}\bigg\{\gamma^{[a}(-\frac{1}{2\alpha}D^{b]}\alpha \lambda_i+\frac{1}{2}D^{b]}\lambda_i-\frac{3i}{8\alpha}G^{b]c}\gamma_c \lambda_i\nonumber\\
					&+\frac{i}{8}T_{ij}\cdot\gamma \gamma^{b]}\lambda^{j})+\gamma^{ab}(\frac{1}{10\alpha}\cancel{D}\alpha\lambda_i-\frac{1}{10}\cancel{D}\lambda_i+\frac{i}{40\alpha}G\cdot \gamma\lambda_i-\frac{i}{40}T_{ij}\cdot\gamma\lambda^{j})\bigg\}\nonumber\\
					&+\frac{i}{48}e_{[\mu}{}^a\bar{\phi}_{\nu]}\gamma^{cd}R(Q)_{cd}-\frac{i}{2}\bar{\phi}_{[\nu}^i\bigg(\frac{i}{4\alpha}G_{ab}\gamma_b\psi_{\mu],i}+\frac{i}{4}T_{cd,ij}\gamma^{cd}\gamma_a\psi_{\mu]}^j+\frac{1}{2}\bar{\psi}_{\mu]}\gamma_a\lambda\lambda_i\bigg)\nonumber\\
					&+\frac{1}{2}\bar{\phi}_{[\nu}\gamma_a\phi_{\mu]}\nonumber\\
			\end{align}}

			\bibliography{references}
			\bibliographystyle{jhep}
		\end{document}